\documentclass{aastex631}

\usepackage{graphicx} 
\usepackage{rotating}
\graphicspath{{figures/}}
\usepackage{graphicx} 
\usepackage{wrapfig}
\usepackage[caption=false]{subfig}

\usepackage[]{hyperref}
\hypersetup{colorlinks = true, linkcolor = blue, anchorcolor = red, citecolor = blue, filecolor = red, urlcolor = red, hypertexnames=true}

\received{\today}
\revised{\today}
\accepted{\today}
\submitjournal{JAAVSO}
\shorttitle{Visual Binaries}
\shortauthors{Soh et al.}

\begin{document}

\title{Markov Chain Monte Carlo Optimization applied to Dyson's Visual Double Stars}
\correspondingauthor{Timothy Banks}
\email{tim.banks@nielsen.com}

\author{Isabella Soh Xiao Si}
\affiliation{Department of Statistics \& Data Science, National University of Singapore,  Blk S16, Level 7, 6 Science Drive 2, Singapore 117546}

\author{Michael D. Rhodes}
\affiliation{Brigham Young University, Provo, Utah 84602, USA}

\author{Edwin Budding}
\affiliation{Carter Observatory, 40 Salamanca Road, Kelburn, Wellington 6012, New Zealand}
\affiliation{School of Chemical \& Physical Sciences, Victoria University of Wellington, PO Box 600, Wellington 6140, New Zealand}

\author{Timothy Banks}
\affiliation{Department of Physical Science \& Engineering, Harper College, 1200 W Algonquin Rd, Palatine, IL 60067, USA}
\affiliation{Data Science, Nielsen, 200 W Jackson, Chicago, IL 60606, USA}


\begin{abstract}
\noindent
Estimates of orbital parameters were made using a Bayesian optimization technique on astrometric data for 25 visual binary systems catalogued a century ago by the ninth Astronomer Royal, Sir Frank Dyson.  An advantage of this method is that it provides reliable, unbiased uncertainty estimates for the optimized parameters. Reasonable agreement is found for the short period ($ < 100$ yr) systems between the current study and Dyson, with superior estimation for the longer systems through the inclusion of an additional century of data. Dynamical masses are presented for the systems through the inclusion of parallax measurements. 

\end{abstract}




\section{Introduction} \label{sec:intro}

Using the classical theory of Keplerian motion of two mass points together with reliable data on their distances and orbital periods, observations of visual binary stars reveal useful physical characteristics of stars.  However, the proportion of visual binaries for which elliptic orbital motion could be clearly established using Earth-based data has been relatively small, involving periods of up to a few hundred years, while  most known visual pairs have orbital periods in the thousands of years. Difficulties in finding accurate distances have also limited the extent to which double star astrometry could significantly bear on stellar astrophysics until recently. Increased precision of double star data including parallaxes, obtained from modern facilities such as the {\em Gaia} mission, is changing our perspective on this. It is appropriate to re-assess procedures for optimal parametrization of visual binary data.  This paper describes the application of statistical optimization techniques to 25 visual binary systems, giving optimal estimates and associated uncertainties for orbital parameters. 

An orbit can be described on the $xy$ plane as (see, e.g. \citealt{Ribas_2002}):
\begin{equation}
    x = \frac{a ( 1 - e^2 )}{1 + e \: cos{\nu}} \big [ \cos{ (\nu + \omega)} \sin{\Omega} + sin{(\nu + \omega)} \cos{\Omega} \cos{i} \big ]
\end{equation}
\begin{equation}
    y = \frac{a ( 1 - e^2 )}{1 + e \: cos{\nu}} \big [ \cos{(\nu + \omega)} \cos{\Omega} - sin{(\nu + \omega)} \sin{\Omega} \cos{i} \big ]
\end{equation}
$a$ is the semi-major axis of the orbit, measured in arc-seconds.  $e$ is the orbital eccentricity. $\nu$ is the true anomaly (or function of time) of the orbit of the stars about their barycenter. $i$ is the inclination, the angle between the plane of projection and the orbital plane. An inclination of 90 degrees would indicate that the orbital plane was exactly side on to our line of sight. $\omega$ is the argument of periastron, being the angle between the node and periastron (closest approach of the two stars). $\Omega$ is the position of the ascending node, which is the position angle of the intersection between the plane of projection and the actual plane the orbit lies in.  These equations were used by the current study as the `model function' for an optimization algorithm. Initially we included two additional parameters $\delta x$ and $\delta y$ to represent offsets in the origin of the $xy$ coordinate system. These will be discussed later in Section~\ref{sec:parameter_reduction}. A second function was used to measure how well the model, given a set of parameters, fitted the data.  This is called the fitting (or optimization) function. The role of the optimization technique was to judiciously adjust estimates for the parameters until a set is reached that well fits the data set. In other words, the optimizer trialed different parameter values in the model function/s, measuring how well the model based on these functions fitted the  data set.  The measure of fit was based on the fitting function. Changes in the parameter estimates led to better or worse fits by the model to the data.  The role of the optimizer is to adjust the parameter estimates until an optimal fit to the data is reached.  In this paper we made use of optimization technique called Markov Chain Monte Carlo, fitting data of 25 double stars.  


\section{MCMC models} \label{sec:mcmc}

A Markov model gives the transition probabilities between one state and another.  A series of such transitions or steps is called a Markov Chain. The key properties of such a process are that it is random and that each state (or step) is independent of the previous step. In other words, the future state of the process depends only on the current state of the process, it does not depend on any past states. 

Markov Chain Monte Carlo (MCMC, see \citealt{Robert_2010} and \citealt{Privault_2013}) models are a Bayesian technique which allow modelling of a distribution, and in particular statistics about that distribution such as the mean or variance. To understand a given distribution, many representative samples are taken from it. Such random samples are called Monte Carlo samples, explaining part of the process name.  The actual distribution itself does not need to be known, all that is required by the technique is to be able to calculate a measure of ``goodness of fit'' of a model to a given data set. MCMC will vary the parameters input into the basic model, leading to model solutions with varying levels of fit to the data. A greater density of such sample points will be in regions where the parameters better fit the data. The simplest MCMC process is the Metropolis algorithm, which is a random walk.  The key attribute of MCMC is that the distribution of interest is sampled again and again by taking small steps across it, building up a map of the distribution. In our case, where we fitted orbital models to observational data, MCMC allowed us to provide statistical estimates of the model parameters and how accurately we can measure those parameters.

\begin{sidewaystable}
    \caption{{\bf Parameter estimates from MCMC fitting} to the Dyson systems, numbered by appearance in Dyson (e.g., 1 refers to Dyson-1 or D1). Uncertainties are single sigma (one standard deviation). See the text for the explanation of the symbols used as the column titles other than `Epoch', which is the time of phase zero for the orbital ephemeris, the orbital period ($P$) in years, and $\sigma$ which is an estimate of the Gaussian noise of the data observations (the `error' in the x and y coordinates). $\sigma$ is a free parameter in the MCMC fits.  `BD' gives an alternative ID for each system, allowing cross referencing }
    \centering
    \hspace{-2.5cm}
    \begin{tabular}{||l|r|r|r|r|r|r|r|r|l||}
    \hline
System & $P$                & $a$                 & $e$               & $\omega$          & $i$             & $\Omega$        & Epoch              & $\sigma$              &  BD  \\
    \hline
1      & $168.28 \pm 0.64$  & $1.0118 \pm 0.0057$ & $0.319 \pm 0.010$ & $ 370.0 \pm 1.6$  & $44.6 \pm 0.6$  & $180.1 \pm 0.1$ & $1955.08 \pm 0.72$  & $ 0.1008 \pm 0.0026$ & BD+22 146 \\
2      & $144.88 \pm 1.89$  & $0.8388 \pm 0.0170$ & $0.230 \pm 0.039$ & $ 364.0 \pm 10.3$ & $62.2 \pm 1.5$  & $100.4 \pm 1.6$ & $1913.08 \pm 3.81$  & $ 0.2191 \pm 0.0083$ & BD+46 536 \\
3      & $872.48 \pm 211.59$& $0.8439 \pm 0.1208$ & $0.825 \pm 0.005$ & $ 282.4 \pm 46.8$ & $157.3 \pm 7.3$ & $28.7 \pm 46.0$ & $1914.73 \pm 1.41$  & $0.1060 \pm 0.0036$  & BD+23 473 \\
4      & $187.04 \pm 4.37$  & $0.5792 \pm 0.0291$ & $0.458 \pm 0.052$ & $ 28.4 \pm 8.5$   & $56.9 \pm 1.6$  & $74.5 \pm 2.7$  & $1886.49 \pm 3.40$  & $ 0.0798 \pm 0.0043$ & BD+31 737 \\
5      & $94.12 \pm 0.37$   & $0.7487 \pm 0.0170$ & $0.597 \pm 0.022$ & $ 303.6 \pm 2.4$  & $53.8 \pm 1.2$  & $143.7 \pm 1.5$ & $1981.09 \pm 0.46$  & $ 0.0821 \pm 0.0041$ & BD+13 728 \\
6      & $57.38 \pm 0.45$   & $0.3413 \pm 0.0340$ & $0.722 \pm 0.066$ & $ 229.5 \pm 5.4$  & $71.4 \pm 2.3$  & $5.3 \pm 2.2$   & $1942.04 \pm 0.56$  & $ 0.0693 \pm 0.0038$ & BD+1 1959 \\
7      & $105.58 \pm 0.55$  & $0.3484 \pm 0.0057$ & $0.413 \pm 0.024$ & $ -16.0 \pm 7.1$  & $27.4 \pm 2.9$  & $145.3 \pm 5.9$ & $1986.60 \pm 0.84$  & $ 0.0632 \pm 0.0026$ & BD+54 1331 \\
8      & $632.95 \pm 221.39$& $1.4929 \pm 0.4947$ & $0.954 \pm 0.024$ & $ 72.4 \pm 40.6$  & $147.1 \pm 19.2$& $150.9 \pm 46.1$& $1894.42 \pm 4.1$   & $0.1067 \pm 0.0038$  & BD+26 2345 \\
9      & $291.69 \pm 19.65$ & $0.8702 \pm 0.0616$ & $0.604 \pm 0.052$ & $ 147.3 \pm 11.4$ & $134.2 \pm 4.6$ & $196.9 \pm 6.5$ & $1872.97 \pm 4.4$   & $0.1505 \pm 0.0049$  & BD+37 2433 \\
10     & $155.95 \pm 0.39$  & $2.4417 \pm 0.0071$ & $0.451 \pm 0.005$ & $ 199.8 \pm 0.8$  & $47.1 \pm 0.3$  & $155.7 \pm 0.4$ & $1916.72 \pm 0.24$  & $ 0.1307 \pm 0.0030$ & BD+27 2296 \\
11     & $273.31 \pm 1.60$  & $0.9240 \pm 0.0195$ & $0.497 \pm 0.037$ & $ 359.0 \pm 1.0$  & $63.8 \pm 1.9$  & $259.0 \pm 1.3$ & $1874.95 \pm 2.12$  & $ 0.0992 \pm 0.0041$ & BD+10 2739 \\
12     & $88.44 \pm 0.45$   & $0.3225 \pm 0.0084$ & $0.551 \pm 0.029$ & $ 131.5 \pm 78.1$ & $170.6 \pm 5.8$ & $37.9 \pm 78.1$ & $1883.65 \pm 0.73$  & $ 0.0687 \pm 0.0029$ & BD+42 2531 \\
13     & $290.86 \pm 4.00$  & $1.4049 \pm 0.0090$ & $0.599 \pm 0.010$ & $ 359.7 \pm 0.3$  & $138.8 \pm 0.9$ & $179.0 \pm 0.8$ & $1872.92 \pm 0.39$  & $ 0.1213 \pm 0.0032$ & BD+37 2636 \\
14     & $219.53 \pm 0.98$  & $2.2300 \pm 0.0209$ & $0.755 \pm 0.007$ & $ 107.6 \pm 1.1$  & $109.9 \pm 0.9$ &  $71.8 \pm 0.03$& $1701.34 \pm 0.10$  & $ 0.3547 \pm 0.0109$ & BD+18 3182 \\
15     & $129.04 \pm 0.53$  & $0.9566 \pm 0.0120$ & $0.618 \pm 0.013$ & $ 145.1 \pm 4.6$  & $25.9 \pm 1.8$  & $62.2 \pm 4.3$  & $1938.77 \pm 0.35$  & $ 0.1422 \pm 0.0036$ & BD+2 3118 \\
16     & $123.27 \pm 0.89$  & $0.9364 \pm 0.0138$ & $0.323 \pm 0.025$ & $ 207.7 \pm 4.7$  & $63.3 \pm 0.9$  & $59.3 \pm 1.0$  & $1893.14 \pm 1.50$  & $ 0.1358 \pm 0.0561$ & BD+43 2639 \\
17     & $265.57 \pm 6.45$  & $1.0063 \pm 0.0141$ & $0.552 \pm 0.019$ & $ 242.9 \pm 4.4$  & $30.4 \pm 1.7$  & $51.3 \pm 3.3$  & $1895.46 \pm 1.04$  & $ 0.1159 \pm 0.0040$ & BD+28 2624 \\
18     & $84.86 \pm 0.94$   & $0.2584 \pm 0.0256$ & $0.650 \pm 0.119$ & $ 12.9 \pm 7.2$   & $125.8 \pm 7.1$ & $152.7 \pm 7.1$ & $2049.18 \pm 5.78$  & $ 0.0545 \pm 0.0045$ & BD+56 1959 \\
19     & $299.9 \pm 24.8$   & $1.1592 \pm 0.0509$ & $0.612 \pm 0.046$ & $ 309.1 \pm 3.8$  & $102.0 \pm 1.0$ & $70.9 \pm 1.0$  & $1915.57 \pm 1.92$  & $ 0.1826 \pm 0.0065$ & BD+03 3610 \\
20     & $347.5 \pm 2.4$    & $1.0369 \pm 0.0279$ & $0.632 \pm 0.004$ & $ 359.3 \pm 0.7$  & $106.2 \pm 2.5$ & $103.3 \pm 1.0$ & $1882.96 \pm 3.94$  & $ 0.1098 \pm 0.0036$ & BD+27 3391 \\
21     & $631.5 \pm 76.7$   & $2.4856 \pm 0.1480$ & $0.468 \pm 0.049$ & $ 168.7 \pm 97.8$ & $170.0 \pm 6.4$ & $180.6 \pm 97.5$& $1866.12 \pm 6.60$  & $ 0.2067 \pm 0.0068$ & BD+44 3234 \\
22     & $167.6 \pm 2.8$    & $0.6367 \pm 0.0056$ & $0.023 \pm 0.019$ & $ 92.3 \pm 100.8$ & $48.0 \pm 1.2$  & $154.0 \pm 1.2$ & $1899.67 \pm 46.67$ & $ 0.0725 \pm 0.0029$ & BD+34 3727 \\
23     & $200.0 \pm 3.0$    & $0.8015 \pm 0.0184$ & $0.485 \pm 0.023$ & $ 49.0 \pm 4.1$   & $64.7 \pm 0.8$  & $175.2 \pm 1.3$ & $1898.68 \pm 1.91$  & $ 0.0823 \pm 0.0037$ & BD--6 5604 \\
24     & $92.1 \pm 0.4$     & $0.7045 \pm 0.0382$ & $0.757 \pm 0.027$ & $ 283.0 \pm 30.8$ & $16.5 \pm 8.6$  & $180.7 \pm 31.1$& $1905.29 \pm 0.35$  & $ 0.1065 \pm 0.0050$ & BD+4 4994 \\
25     & $251.3 \pm 29.6$   & $0.9086 \pm 0.0571$ & $0.594 \pm 0.076$ & $ 213.6\pm 13.9$  & $131.1 \pm 6.4$ & $143.3 \pm 7.1$ & $1899.70 \pm 4.5$   & $ 0.0730 \pm 0.0045$ & BD+38 5112 \\
    \hline 
    \end{tabular} 
    \label{tab:dyson_parameters}
\end{sidewaystable}

We used the {\em rstan} library \citep{Stan_2021} for the MCMC modelling  in the R statistical programming language \citep{R_Core_2021}. This library is an interface to the STAN\footnote{See https://mc-stan.org/ for documentation.} programming language. STAN is a C++ library allowing Bayesian inference using the No-U-Turn (NUTS) sampler (a variant of Hamiltonian Monte Carlo (HMC), see \citealt{Hoffman_2014}) or frequentist inference via optimization methods.  The HMC algorithm avoids the random walk and associated sensitivity to correlated parameters experienced by earlier MCMC methods such as the Metropolis or Gibbs samplers.  It does this via examination of first-order gradients to guide its steps.  This improvement leads to more rapid convergence than the previously mentioned methods.  HMC still suffers from high sensitivity to the step size and the number of steps required to reach convergence, which are both user-set parameters.  If these are not set correctly (in particular the step size), HMC will either revert to random walk behavior (when the step size is too small) or waste computation (if the step size is too large). NUTS is a refinement to HMC which removes the need to set a number of steps through use of a recursive algorithm which scans a wide range of possible steps.  The method automatically stops its steps when the chain has doubled back on itself.  \cite{Hoffman_2014} show that NUTS performs at least as efficiently as a well-tuned HMC method, with the advantage of requiring less user input. Given its advantages over earlier MCMC techniques, and the ease to implement the model inside R and STAN, we made use of this technique. 

We are not the first authors to apply a MCMC method to visual double star data, although the technique is not yet widely used in the field.  \cite{Mendez_2017} modeled the orbits of 18 visual binaries using the Differential Evolution MCMC technique, which ran multiple markov chains simultaneously with sharing of information between the chains to aid convergence. \cite{Sahlman_2013} used MCMC to estimate the orbital parameters of a low-mass companion to an ultracool dwarf star.  \cite{Lucy_2014} successfully used a MCMC model to explore numerical simulations modelling total masses of visual binaries with measured parallaxes but incomplete orbits, finding that the mass estimates were unbiased when more than 40\% of the orbit was covered by the data.  \cite{Claveria_2019} applied MCMC to examine the impact (estimating orbital parameters) when partial measurements were included into a data set, finding that such inclusions could lead to more accurate estimation of the parameters in some circumstances.  


\begin{figure}
    \centering 
\begin{subfloat}[Dyson-6]{
    \includegraphics[width=0.45\linewidth]{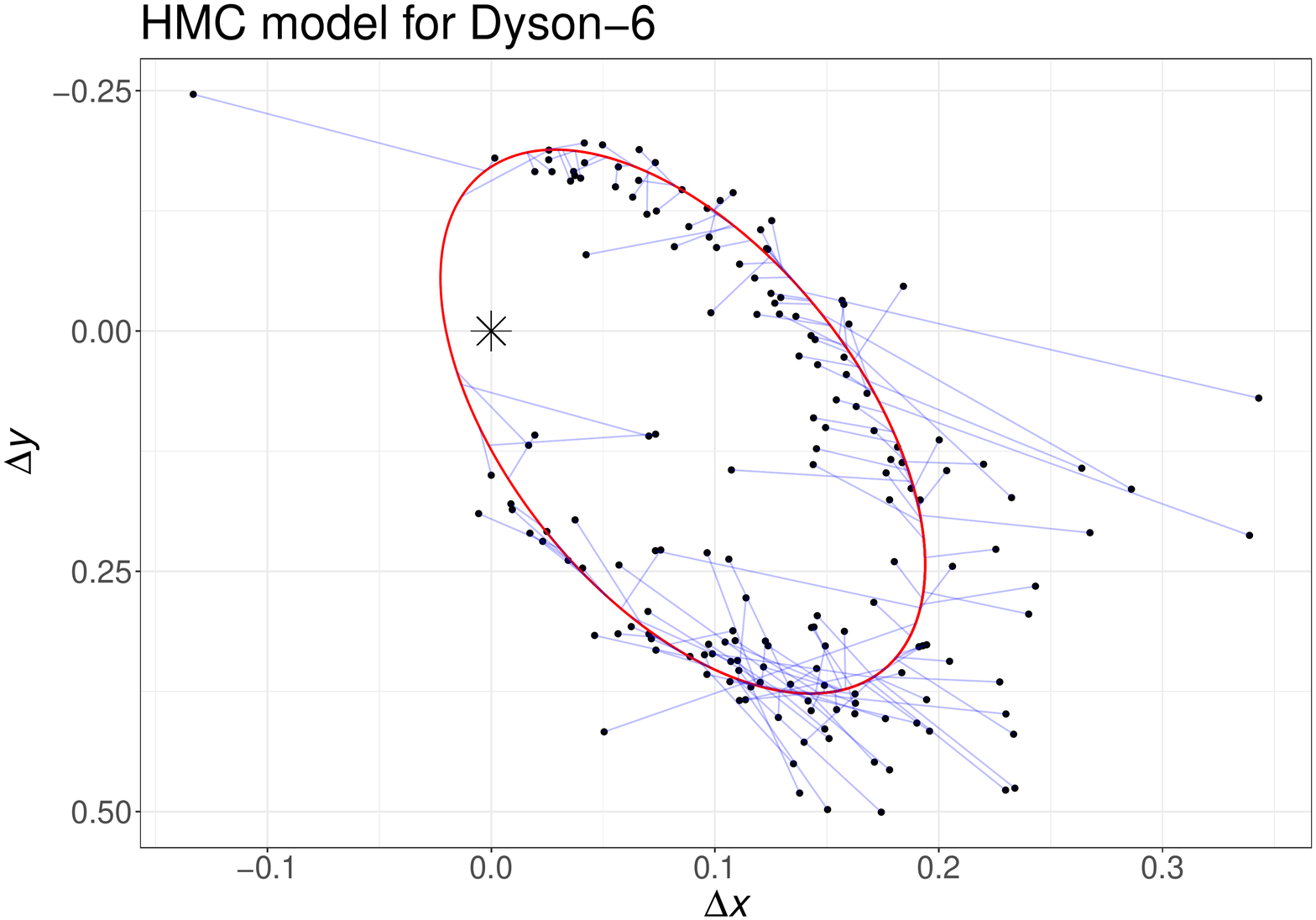} 
    \label{fig:d16_orbit}}
\end{subfloat}\hfil 
\begin{subfloat}[Dyson-10]{
    \includegraphics[width=0.45\linewidth]{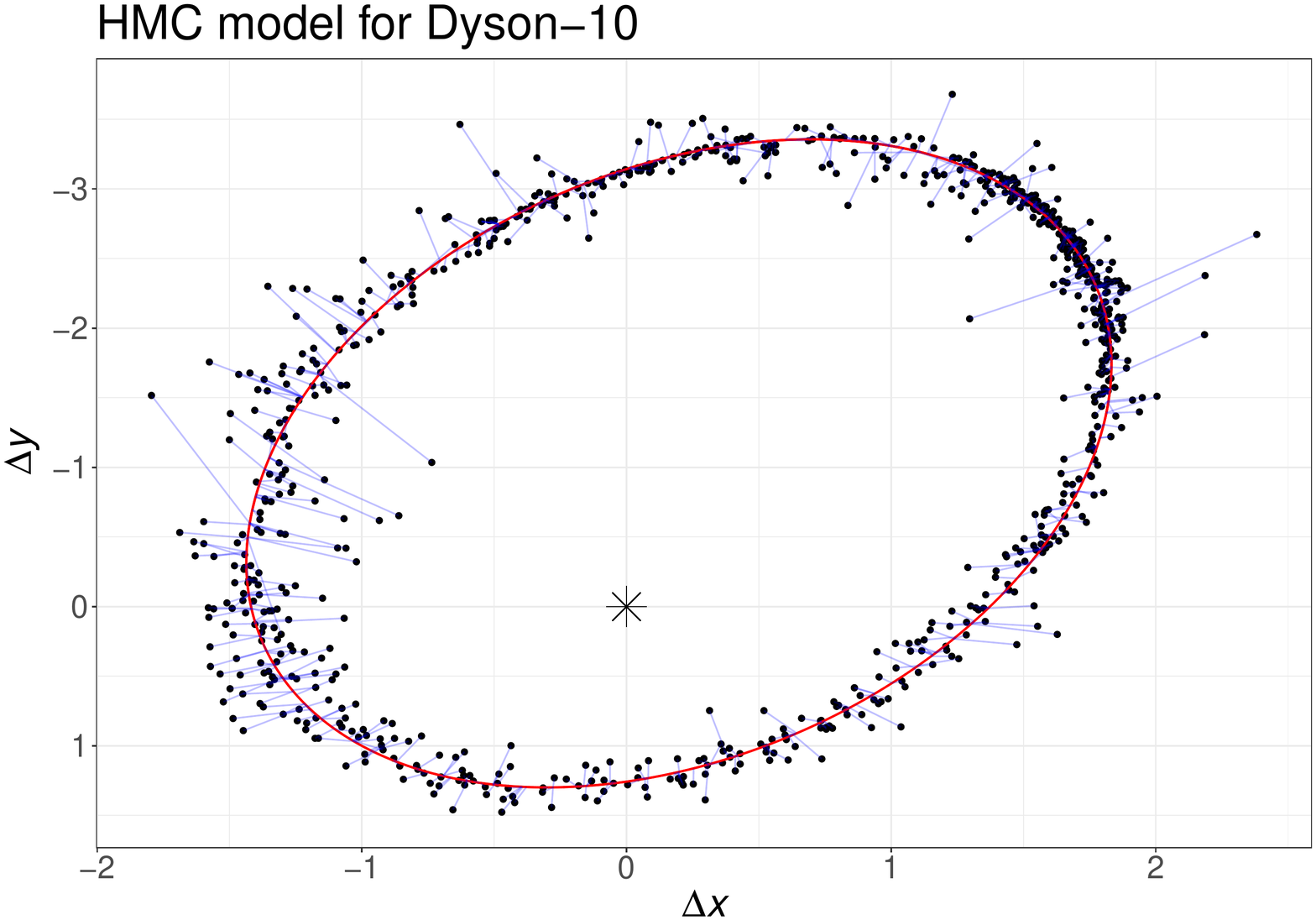} 
    \label{fig:d16_orbit}}
\end{subfloat}\hfil 
\begin{subfloat}[Dyson-18]{
    \includegraphics[width=0.45\linewidth]{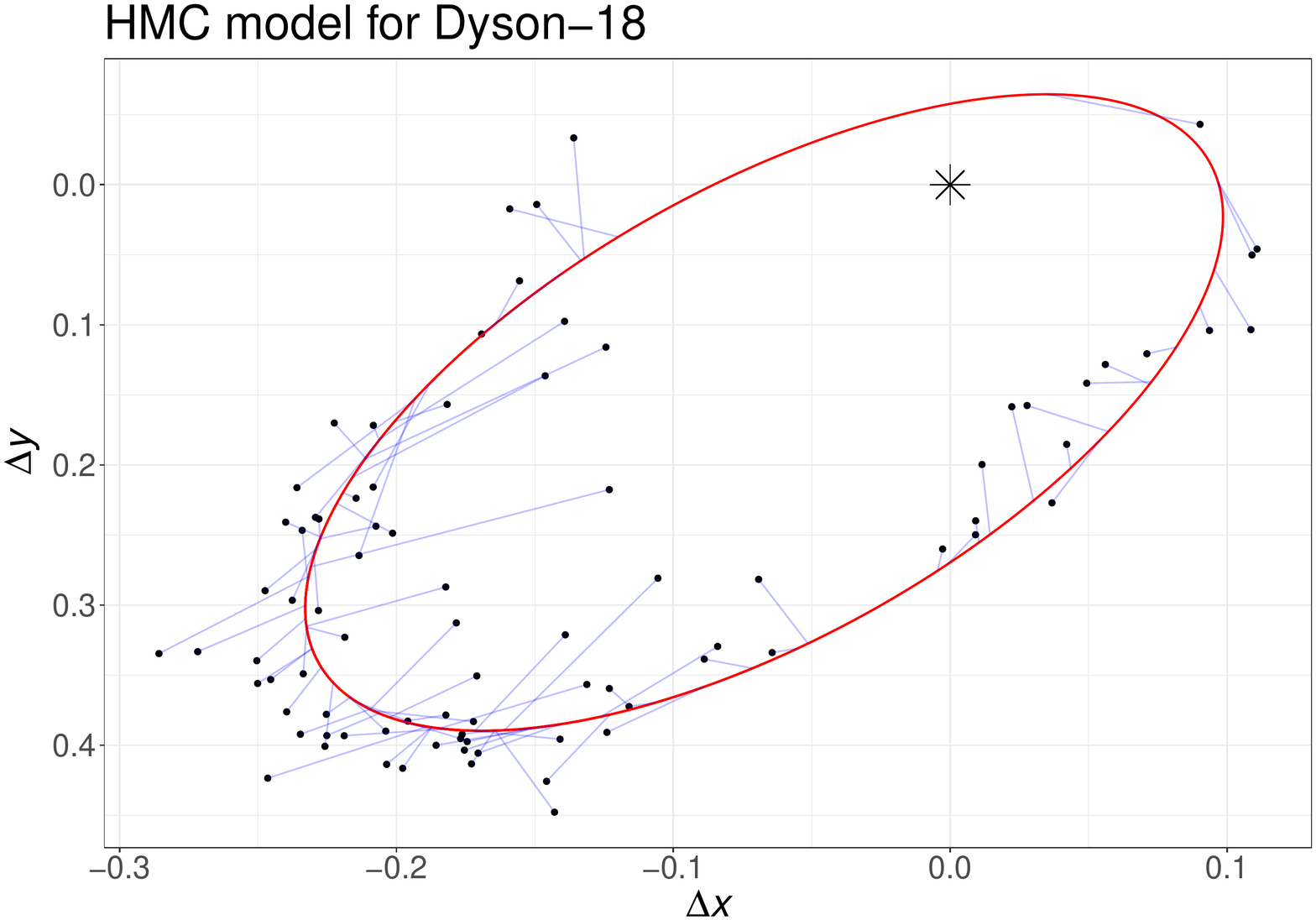} 
    \label{fig:d16_orbit}}
\end{subfloat}\hfil 
\begin{subfloat}[Dyson-21]{
    \includegraphics[width=0.45\linewidth]{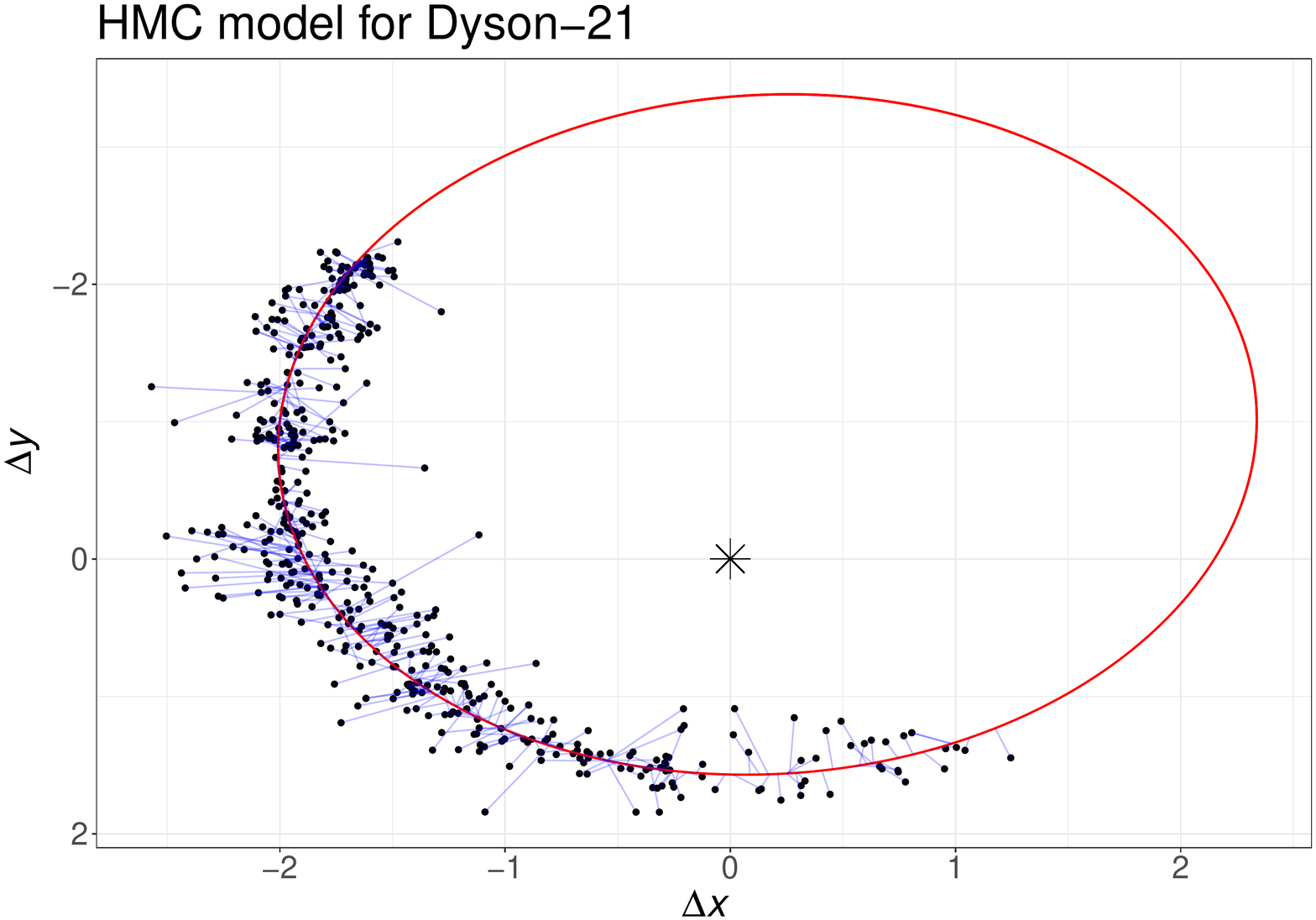} 
    \label{fig:d16_orbit}}
\end{subfloat}\hfil 
\caption{Observations and model orbits for representative Dyson systems selected to show a range of data sets, from complete orbits to partial and of different noise levels.  East increases to the right, and down is northwards (as per convention in many visual binary papers).  The model orbits are shown as the red curve, observations are plotted as black dots, and the green lines connect the observations to their modelled positions on the model orbits. The star symbol at $(\Delta x, \Delta y) = (0,0)$ is the location of the primary star in each system. These parameters should not be confused with $\delta x$ and $\delta y$ mentioned in the paper text, which are adjustments to the origin used by the optimizer to improve the fit. Orbital parameters are given in Table~\ref{tab:dyson_parameters}.
\label{fig:dyson_orbit_models}}
\end{figure}


\section{Study Rationale}

The current paper examines the orbits of 25 visual binaries catalogued by the ninth Astronomer Royal, Sir Frank Dyson.  \cite{Dyson_1921} listed observations spanning from the early nineteenth century to his time of publication.  We have supplemented the data set with further observations collected over the century since 1921, sourced from the Washington Double Star (WDS) catalog \citep{Mason_2022}. \cite{Rhodes_2023} applied a modified version of the Levenberg-Marquardt (see \citealt{Bevington_1969}) optimization technique to this data set, coupled with examination of the $\chi^2$ Hessian matrix (for further details see \citealt{Banks_1990}).  They provided estimates and accompanying uncertainties in the orbital parameters. Agreement between the published WDS estimates and those of \cite{Rhodes_2023} was good. Pending issues from the paper included concerns whether the global minima has been reached by the optimization methods and that they were not local minima `trapping' the search methods. \cite{Rhodes_2023} noted that final solutions were, for some systems, dependent on the starting estimates of the parameters --- suggesting the presence of local minima. There was also interest in using an alternative technique to explore the uncertainties in the parameter estimates, perhaps allowing tuning of the step sizes applied to the curvatures from the Hessian matrix that were used to provide uncertainties for the optimized parameters.  While a grid search might have given insight into the first concern, MCMC could address both and therefore led to this current study.

\begin{sidewaysfigure}  
\centerline{\includegraphics[height=0.65\textheight]{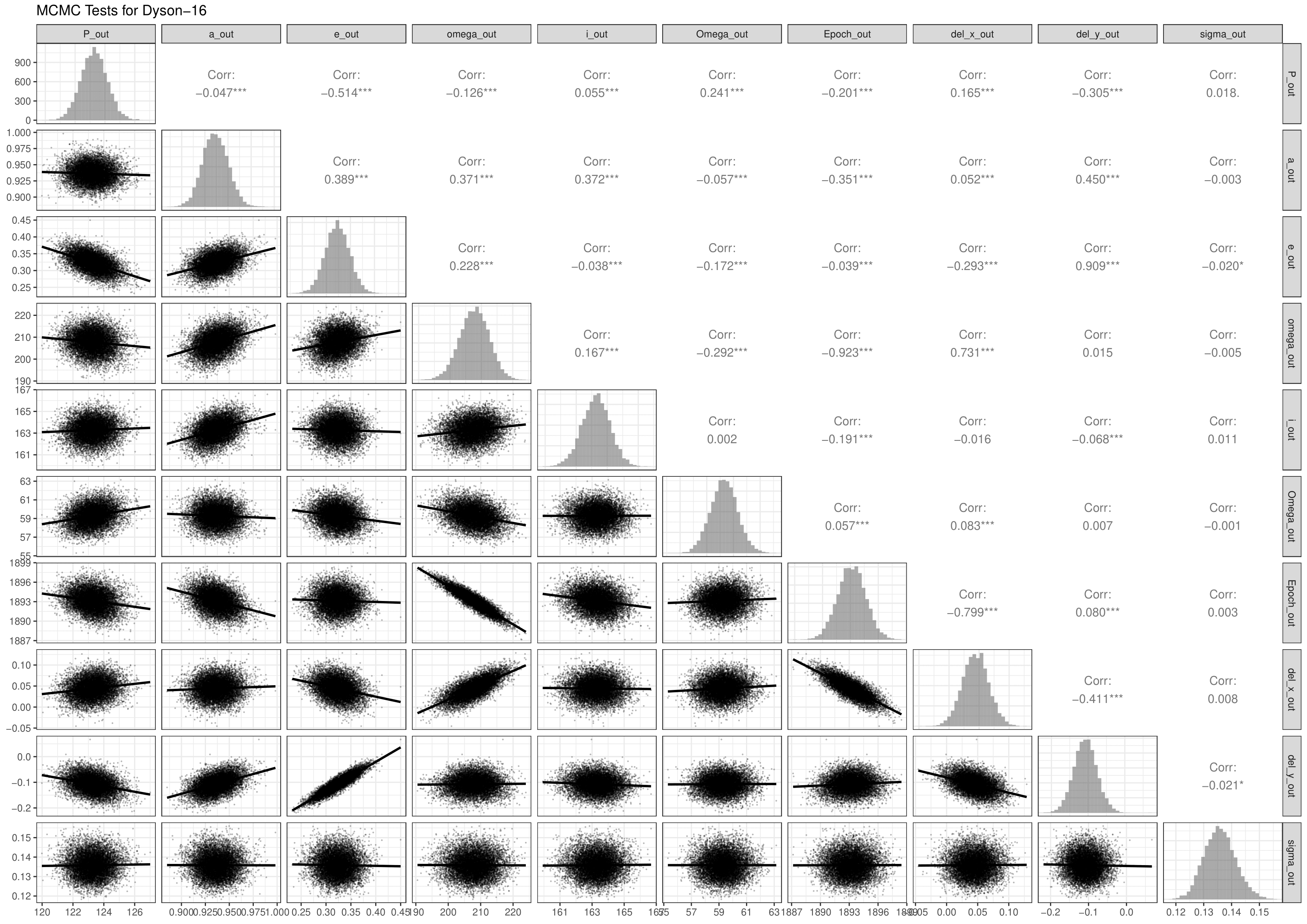}}
\caption{
Example `corner plot' based on the MCMC fitting for Dyson-16, which is representative of corner plots for the other systems.  This represents 10,000 steps in the Markov chain, excluding the initial 10,000 steps known as `burn-in'.  These steps are excluded from the final results, and are considered a result of starting the optimization in a lower probability set of parameters, leading to movement to the global minimum. The columns and rows correspond to the optimized parameters, namely $P$, $a$, $e$, $\omega$, $i$, $\Omega$, epoch, $\Delta x$, $\Delta y$, and $\sigma$, in that order.  The diagonal shows histograms of the parameter estimates, the upper right of the figure gives the correlation coefficients between pairs of the parameters, and the lower left plots the parameters estimates for pairs of the parameters. See text for more information.
\label{fig:example_corner_plot}  
}
\end{sidewaysfigure}


\begin{wraptable}{l}{6.5cm}
\caption{ Dynamical masses (solar units) based on {\em Hipparcos} and {\em Gaia} DR3 parallaxes (where available). Dyson numbers are used to identify the systems. Errors are one standard deviation.}\label{tab:dynamical masses}
\begin{tabular}{ccc}\\ 
\hline
Dyson ID    & Hipparcos       & Gaia \\
\hline
1           & 2.00 $\pm$ 0.04 & \\
2	        & 1.74 $\pm$ 0.06 & \\
3	        & 1.83 $\pm$ 0.52 & \\ 
4	        & 5.81 $\pm$ 0.31 & \\
5	        & 3.20 $\pm$ 0.12 & 4.07 $\pm$ 0.05 \\
6	        & 2.87 $\pm$ 0.19 & \\
7	        & 14.4 $\pm$ 0.16 & \\
8	        & 6.11 $\pm$ 0.76 & 5.05 $\pm$ 0.76 \\
9	        & 1.75 $\pm$ 0.17 & 1.33 $\pm$ 0.16 \\
10	        & 1.44 $\pm$ 0.02 & \\
11	        & 0.85 $\pm$ 0.10 & 0.89 $\pm$ 0.04 \\
12	        & 2.53 $\pm$ 0.12 & \\
13	        & 1.37 $\pm$ 0.05 & 1.75 $\pm$ 0.06 \\
14	        & 1.75 $\pm$ 0.03 & \\
15	        & 7.86 $\pm$ 0.06 & \\
16	        & 1.07 $\pm$ 0.08 & \\
17	        & 2.88 $\pm$ 0.07 & 3.71 $\pm$ 0.04 \\
18	        & 1.54 $\pm$ 0.20 & \\
19	        & 2.85 $\pm$ 0.15 & \\
20	        & 2.48 $\pm$ 0.15 & 2.13 $\pm$ 0.05 \\
21	        & 4.98 $\pm$ 0.20 & 4.07 $\pm$ 0.20 \\
22	        & 2.37 $\pm$ 0.08 & \\
23	        & 2.88 $\pm$ 0.08 & 2.86 $\pm$ 0.05 \\
24	        & 1.36 $\pm$ 0.13 & \\
25	        & 1.45 $\pm$ 0.21 & 1.48 $\pm$ 0.20 \\
\hline
\end{tabular}
\end{wraptable} 

\section{Analysis}

Table~\ref{tab:dyson_parameters} (on page~\pageref{tab:dyson_parameters}) lists the optimal parameter values and associated standard deviations for each of the Dyson systems.  The parameters $\delta x$ and $\delta y$ are not included in the table for reasons of space.  These are zero point adjustments to the co-ordinate system, and fell within one standard deviation of zero in all cases.  Position angles had been precessed to the year 2000 (see page 73 of \citeauthor{Aitken_1964}, \citeyear{Aitken_1964}; page 121 of \citeauthor{Cocteau_1981}, \citeyear{Cocteau_1981}; and page 276 of \citeauthor{Greaney_2004}, \citeyear{Greaney_2004}). This was important for the earlier observations which were collected nearly 175 years before the chosen epoch.

Runs continued until the \^{R} diagnostics (see \citealt{Sinharay_2003}) were within a thousandth of unity, which was typically achieved in 20,000 steps. Figure~\ref{fig:dyson_orbit_models} (on page \pageref{fig:dyson_orbit_models}) plots data for four representative Dyson systems, along with the model orbit based on the optimal parameter values given in table~\ref{tab:dyson_parameters}. Again, not all fits to the data are shown in this paper to conserve space.  

Figure~\ref{fig:example_corner_plot} (on page \pageref{fig:example_corner_plot}) is an example of a `corner plot', which were generated for all fits. This is one of the standard diagnostic tools used in MCMC analysis. This figure is representative of the corner plots for the other systems. The diagonal running from upper left to lower right plots histograms plotting parameter estimates from each of the last 10 thousand MCMC steps, for each of the parameters in turn.  These are all nicely Gaussian shaped in the diagram, indicating a stable solution has been reached. Correlations between the parameters are given in the boxes to the upper right, while the charts to the lower left plot each parameter against the others.  These charts clearly show the correlations between parameters, or lack there of.  

Most systems were simple convex optimizations. D8 (D for Dyson), D12, D18, D20, D21, D22, D24, and D25 were more challenging systems.  Convergence was more difficult due to higher levels of noise and/or shorter orbital arcs being observed for these systems compared to the others in this set. D12, D20, and D21 exhibited double peaks in the posterior distributions, indicating problems with the symmetry of some of the angle parameters.  A point optimization technique could settle in one of these two minima depending on the starting parameter values chosen.

The dynamical (or combined stellar) mass $M_d$ of such binary systems can be calculated if the parallax is known, via an equation \citep{Malkov_2012} based on Kepler's third law:
\begin{equation}
    M_{d} = \frac{a^3}{\pi^3 P^2}
    \label{eq:masses}
\end{equation}
where both $a$ and the parallax $\pi$ are in milli-arcseconds, $P$ is in years, and $M_d$ is in solar masses. Estimated dynamical masses are given in Table~\ref{tab:dynamical masses}, based on {\em Hipparcos} \citep{ESA_1997, Perryman_1997} and {\em Gaia} Data Release 3 \citep{Gaia_2022} parallaxes. Errors in $a$, $P$, and $\pi$ were propagated through equation~\ref{eq:masses} to give the single sigma uncertainties presented in Table~\ref{tab:dynamical masses}. Not all of the systems with parallaxes available from both catalogs have dynamical mass estimates within formal statistical agreement at two sigma.


\begin{figure}
    \centering 
\begin{subfloat}[$P$]{
   \includegraphics[width=0.35\linewidth]{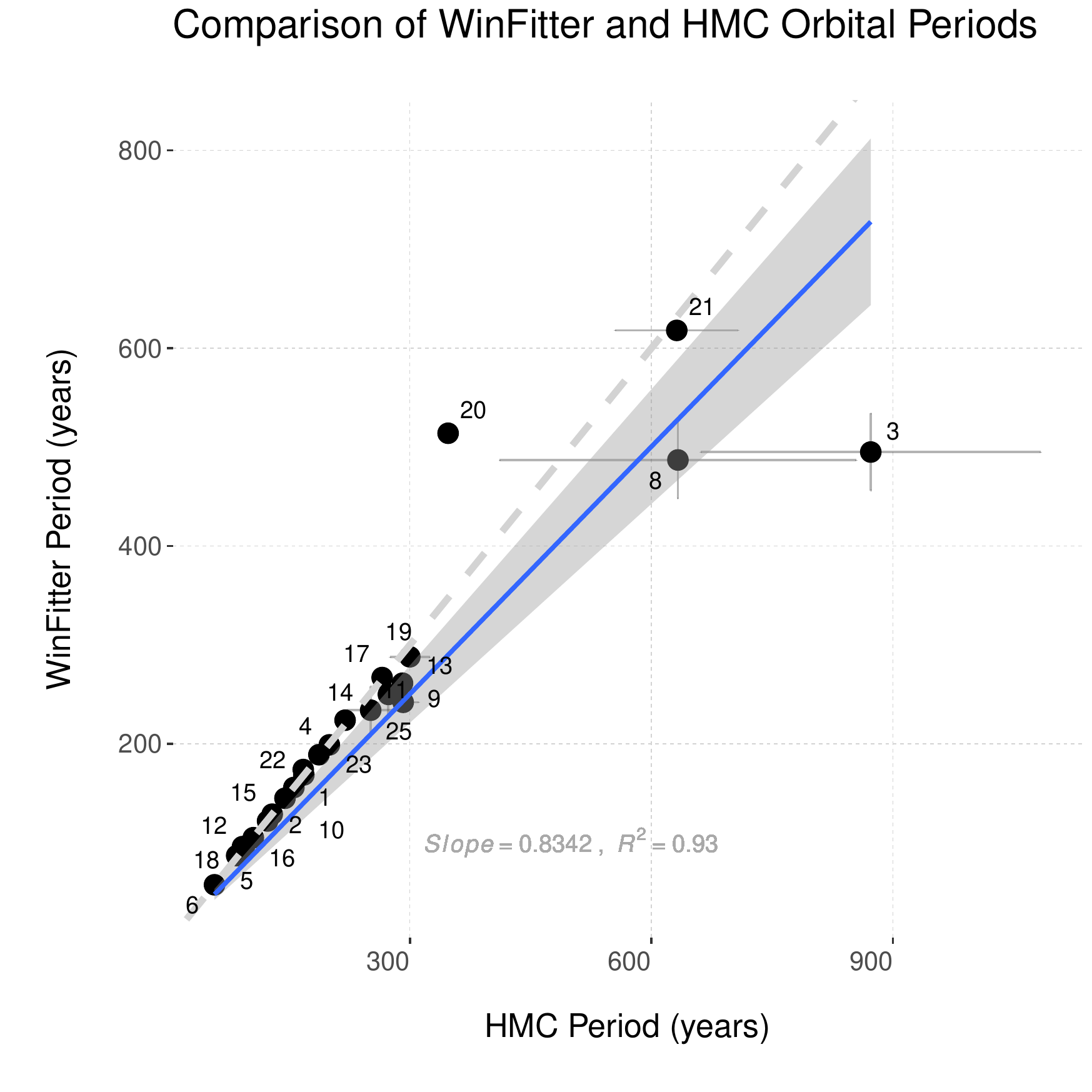} 
   \label{fig:Period_comp}}
\end{subfloat}\hfil 
\begin{subfloat}[$a$]{
   \includegraphics[width=0.35\linewidth]{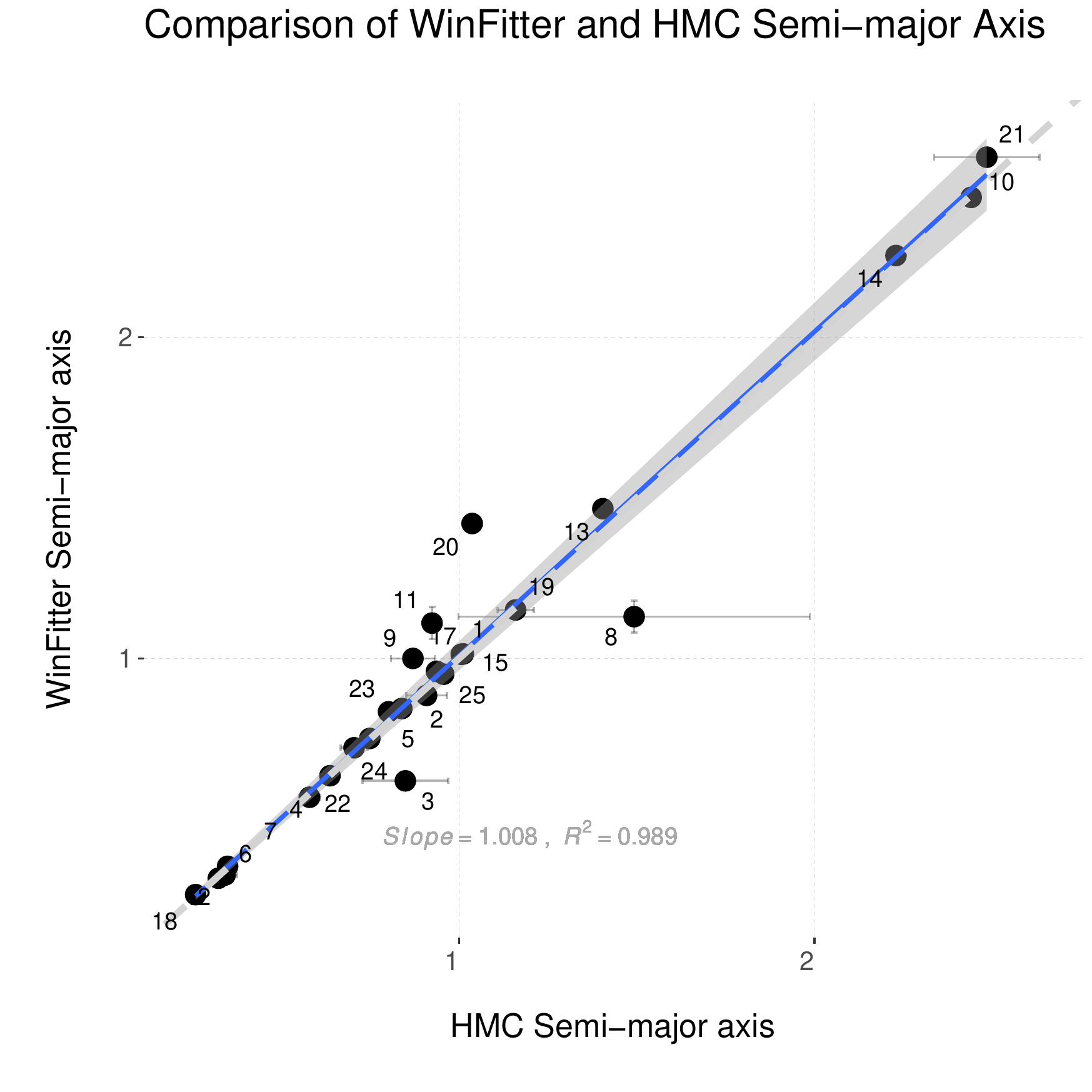} 
   \label{fig:axis_comp}}
\end{subfloat}\hfil 
\begin{subfloat}[$e$]{
   \includegraphics[width=0.35\linewidth]{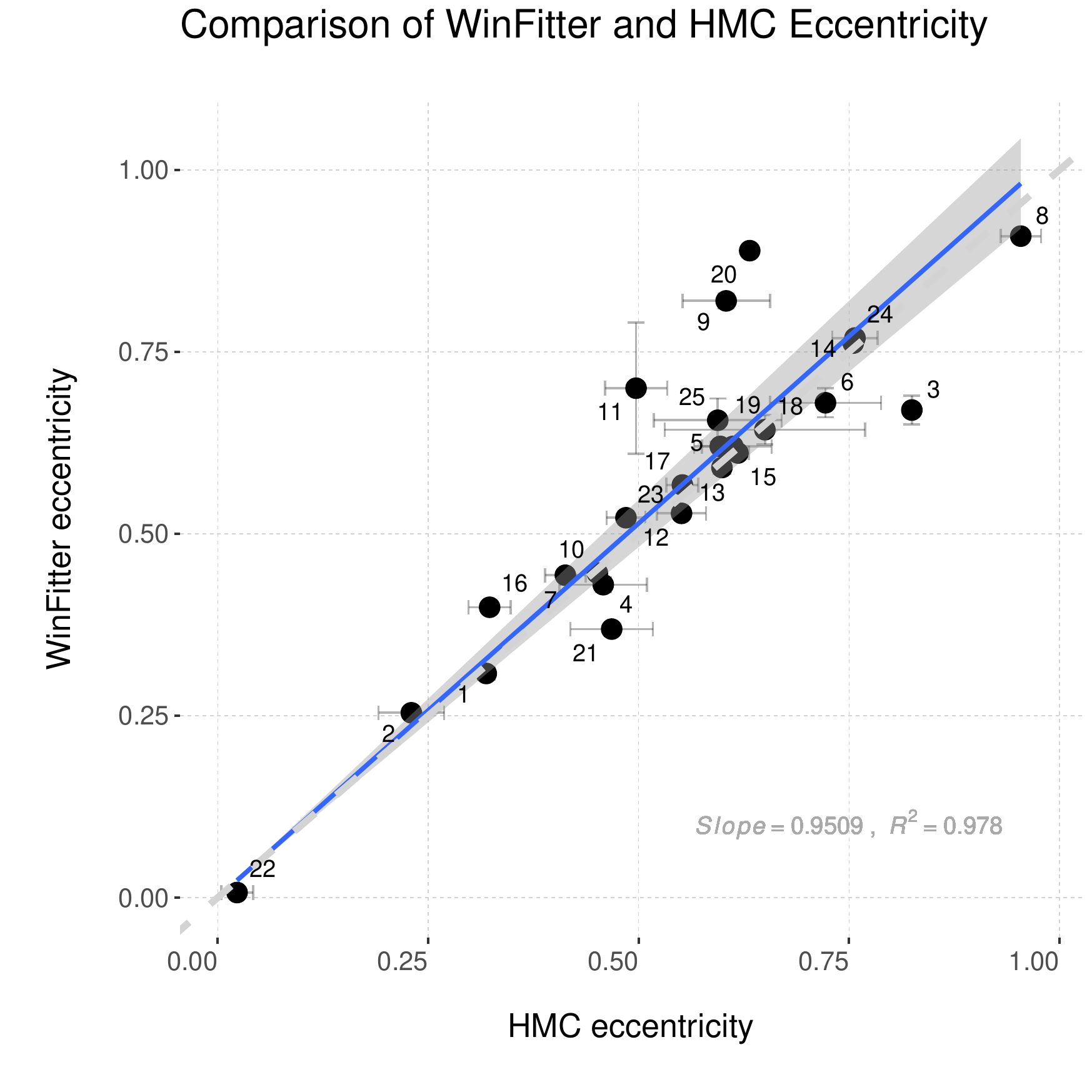} 
   \label{fig:ellip_comp}}
\end{subfloat}\hfil 
\begin{subfloat}[$\omega$]{
   \includegraphics[width=0.35\linewidth]{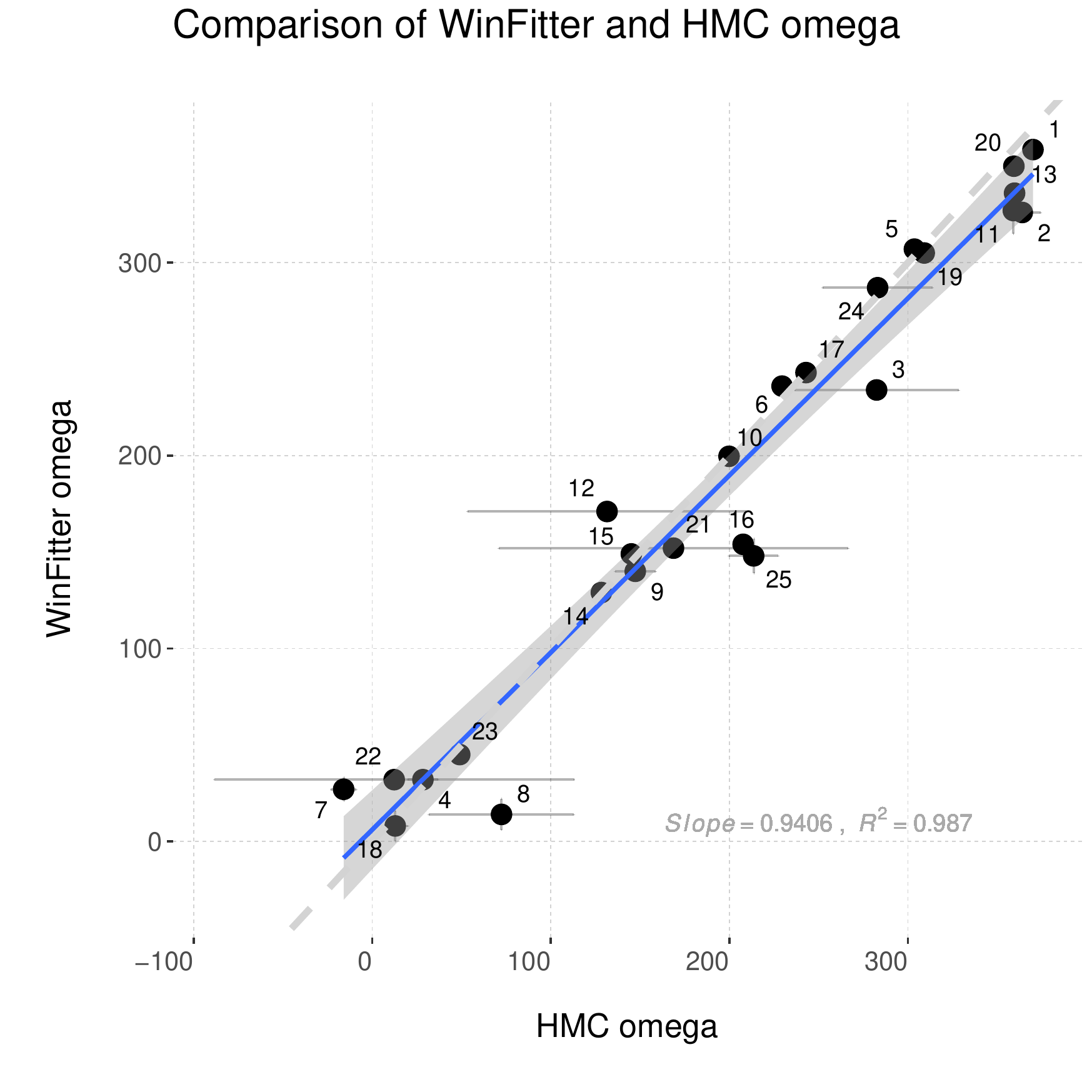} 
   \label{fig:small_omega_comp}}
\end{subfloat}\hfil 
\begin{subfloat}[$i$]{
   \includegraphics[width=0.35\linewidth]{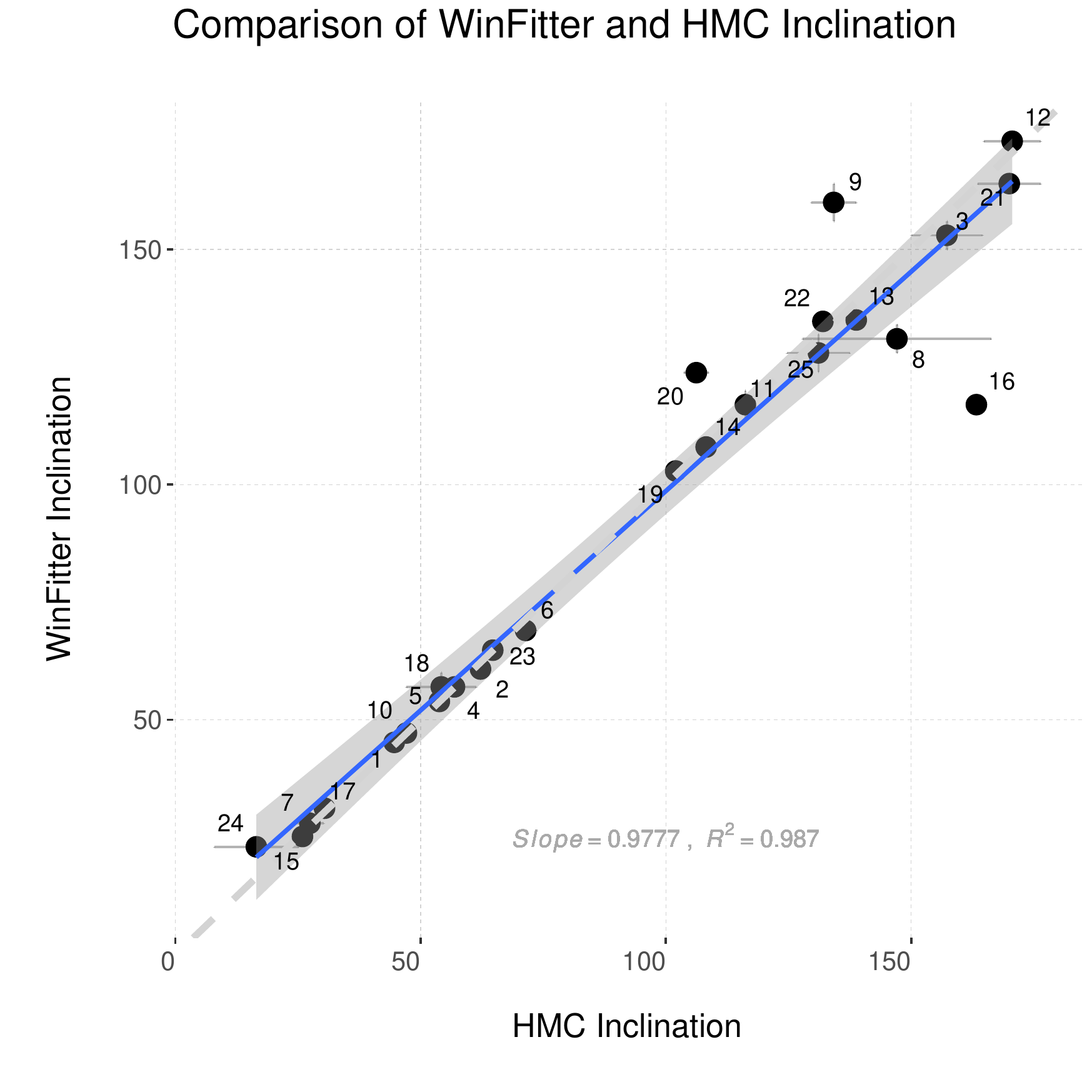} 
   \label{fig:inc_comp}}
\end{subfloat}\hfil 
\begin{subfloat}[$\Omega$]{
   \includegraphics[width=0.35\linewidth]{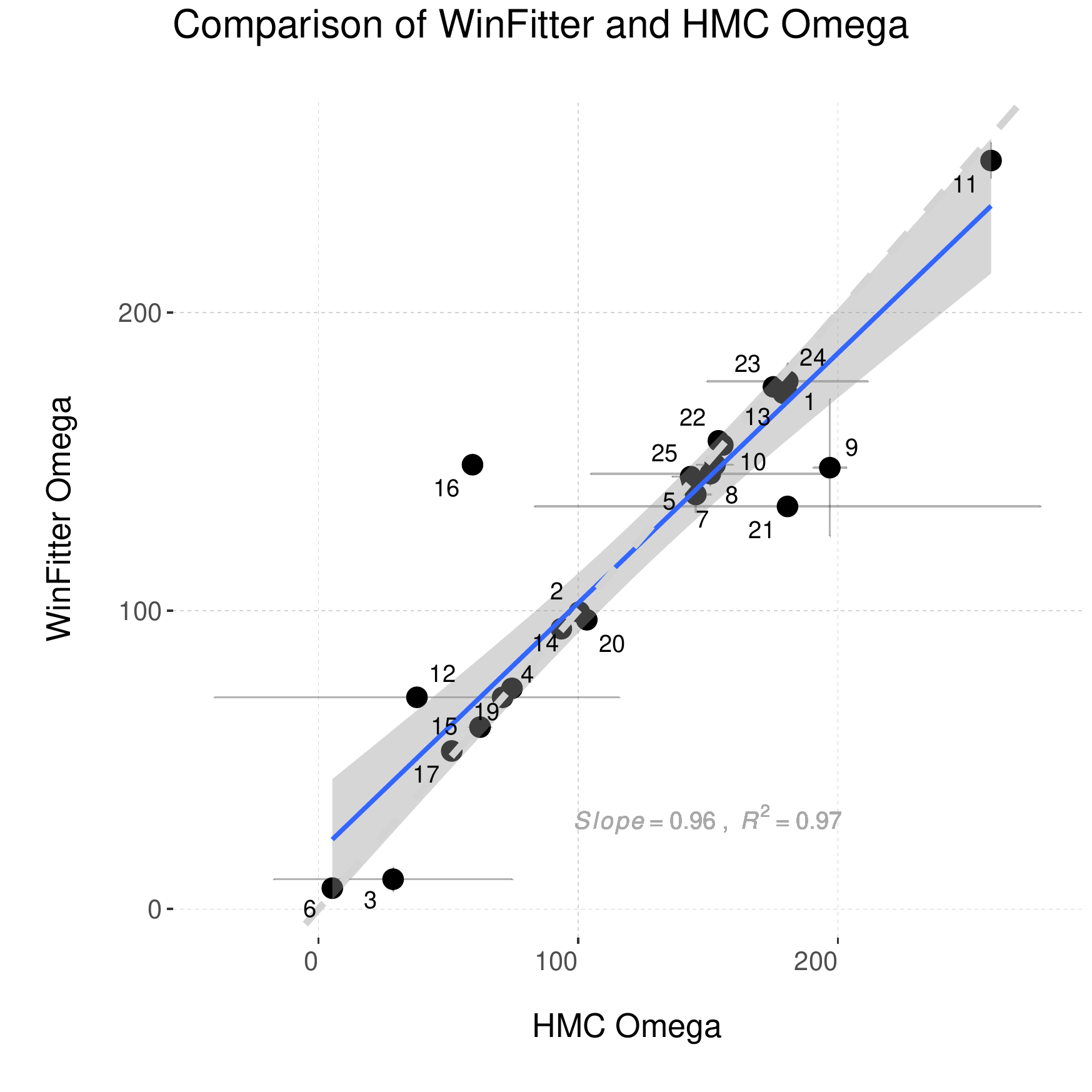} 
   \label{fig:big_omega_comp}}
\end{subfloat}\hfil 

\caption{Comparison between HMC and WinFitter optimal parameter estimates. Systems are denoted by their Dyson number. Regressions have been fitted to the data, resulting in best-fit (blue-colored) lines in the charts. Two sigma confidence limits are shown as the grey shaded regions. The dotted lines are those of perfect agreement.
\label{fig:hmc_wf_comp}}
\end{figure}

\begin{figure}
    \centering 
\begin{subfloat}[$P$]{
   \includegraphics[width=0.35\linewidth]{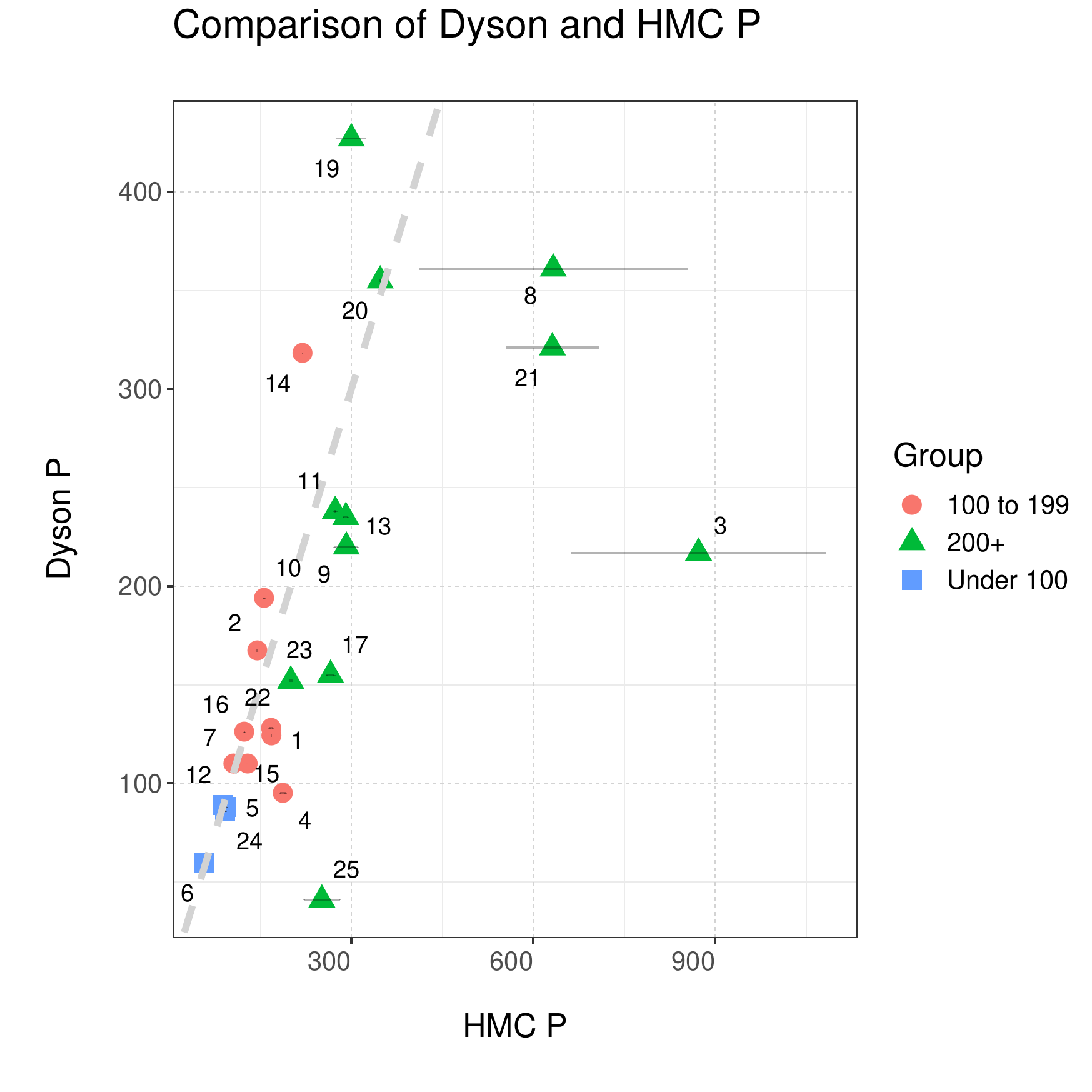} 
   \label{fig:Dyson_P}}
\end{subfloat}\hfil 
\begin{subfloat}[$a$]{
   \includegraphics[width=0.35\linewidth]{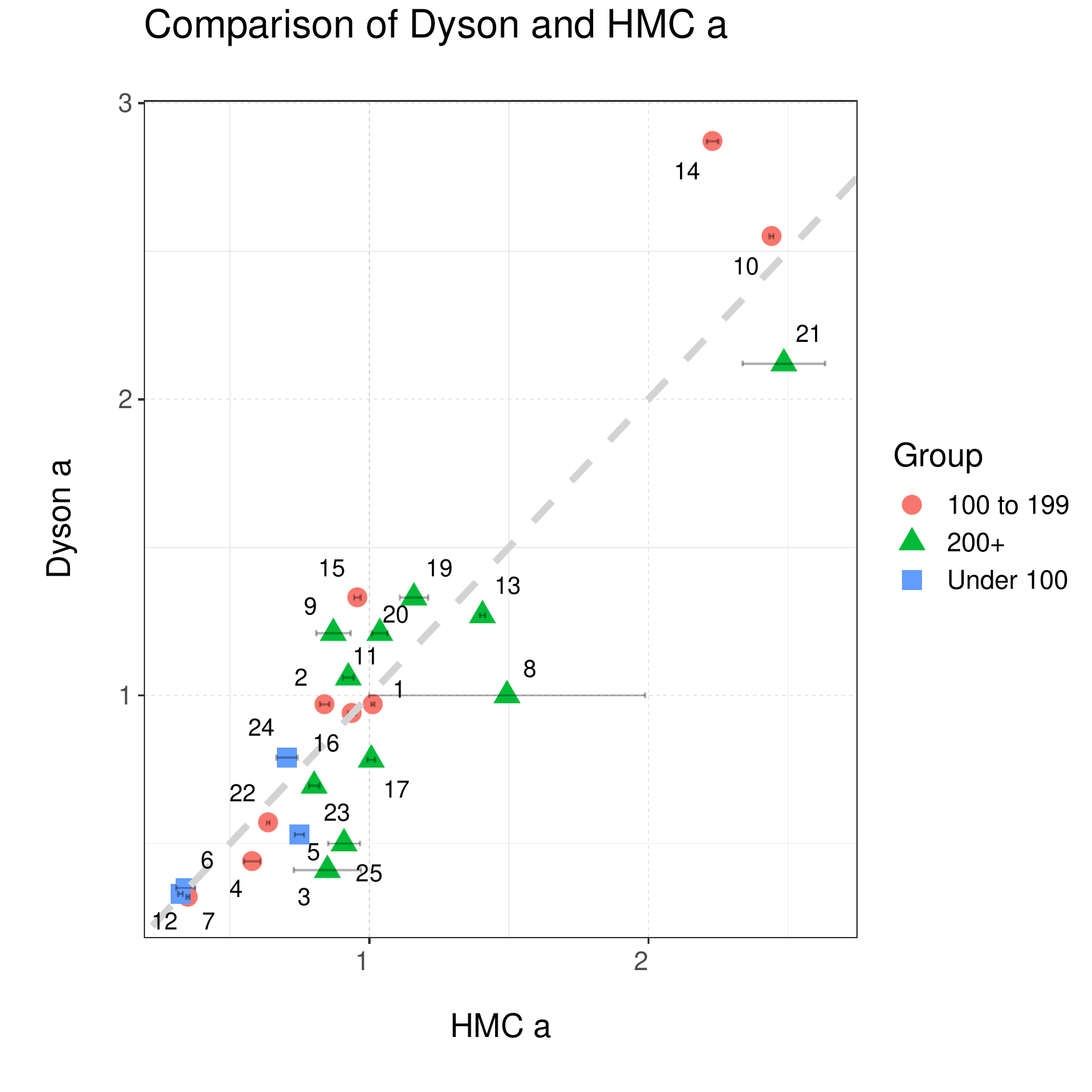} 
   \label{fig:Dyson_a}}
\end{subfloat}\hfil 
\begin{subfloat}[$e$]{
   \includegraphics[width=0.35\linewidth]{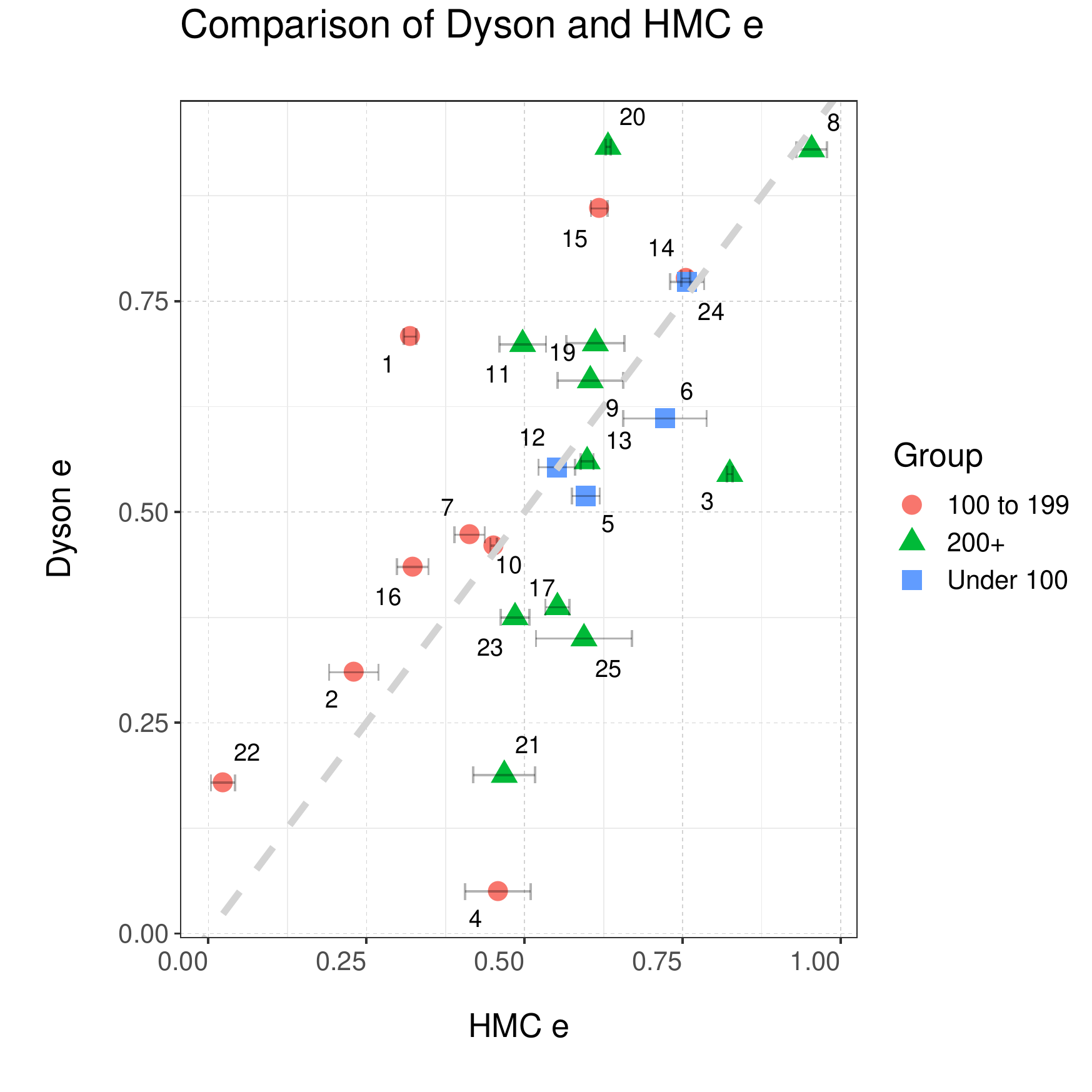} 
   \label{fig:Dyson_e}}
\end{subfloat}\hfil 
\begin{subfloat}[$\omega$]{
   \includegraphics[width=0.35\linewidth]{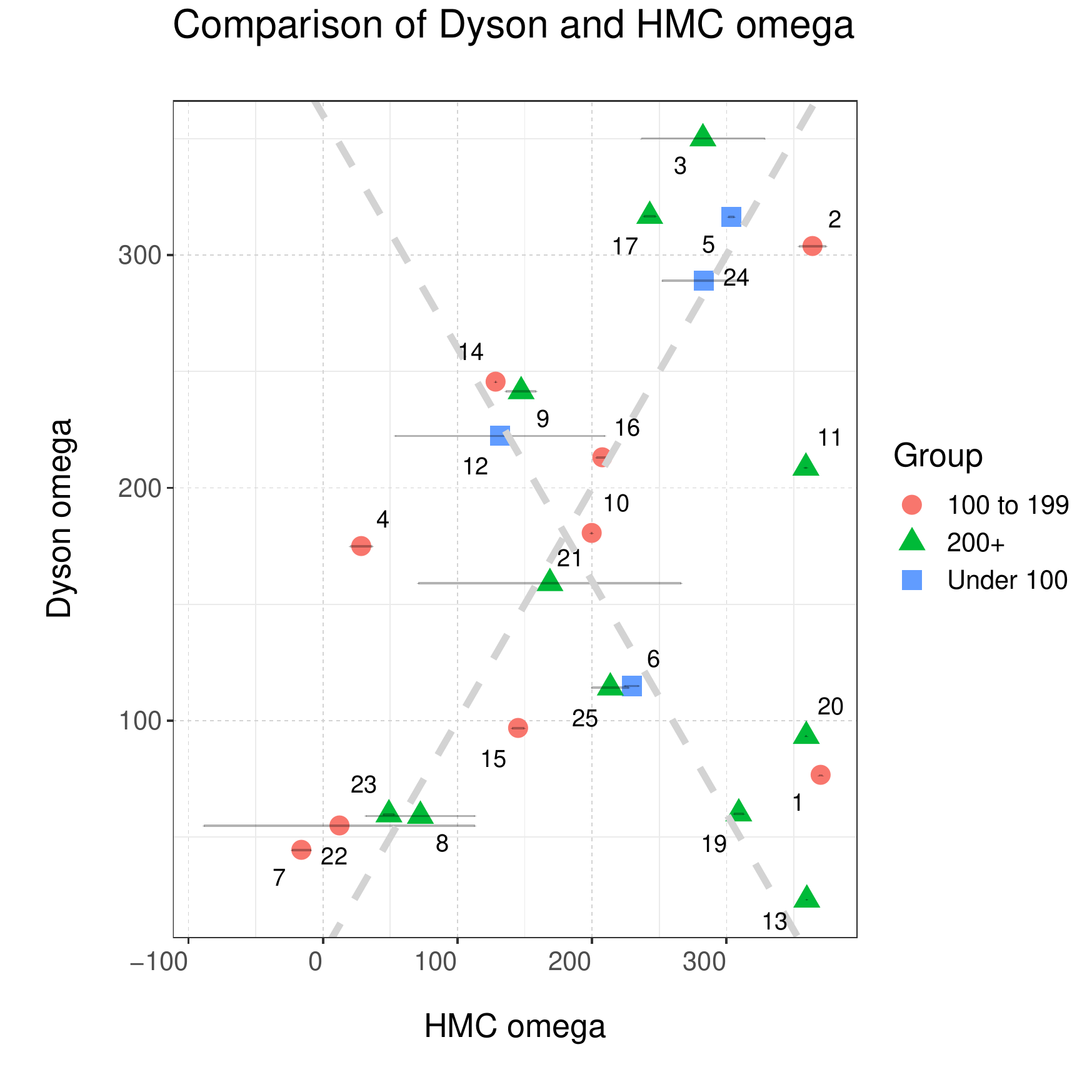} 
   \label{fig:Dyson_omega}}
\end{subfloat}\hfil 
\begin{subfloat}[$i$]{
   \includegraphics[width=0.35\linewidth]{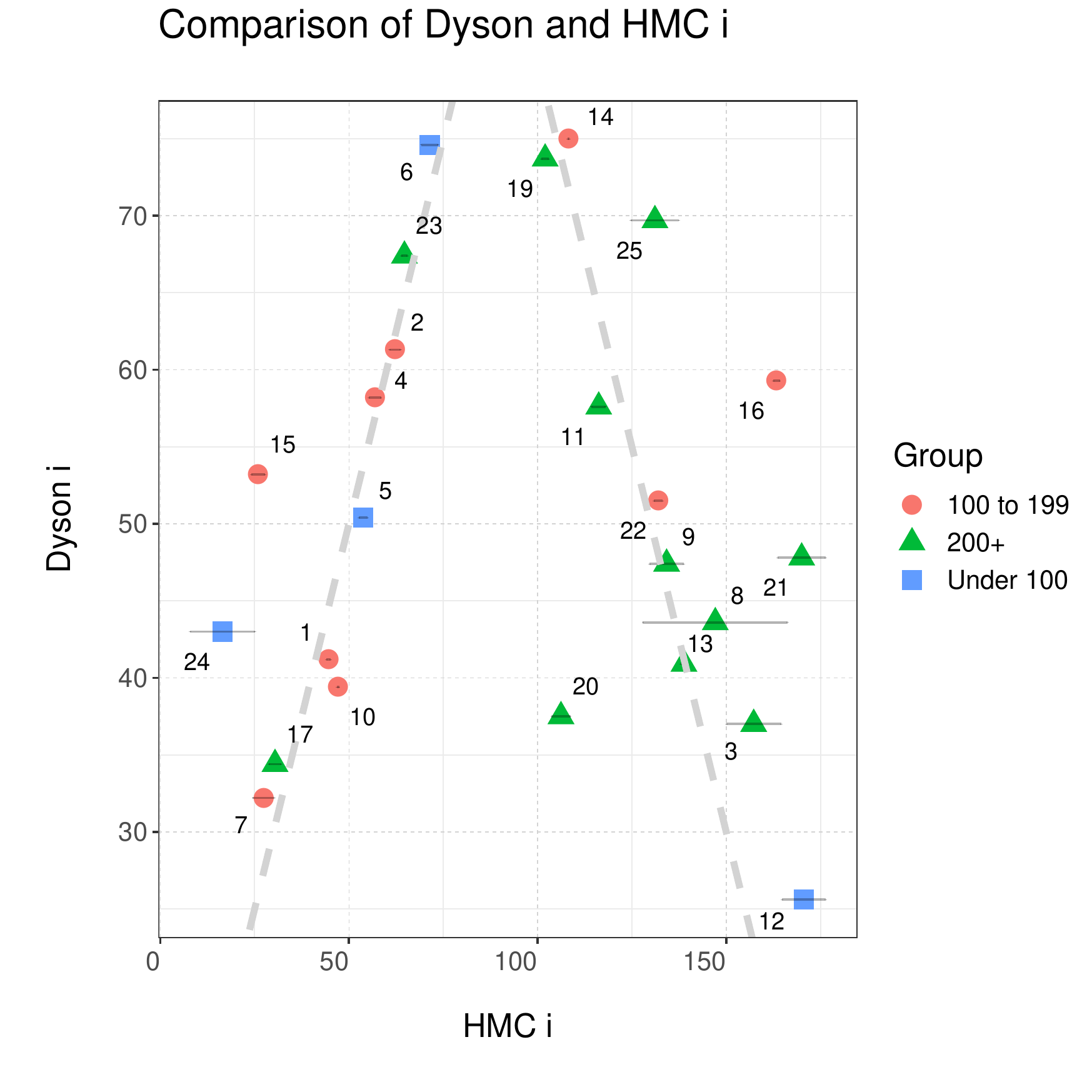} 
   \label{fig:Dyson_i}}
\end{subfloat}\hfil 
\begin{subfloat}[$\Omega$]{
   \includegraphics[width=0.35\linewidth]{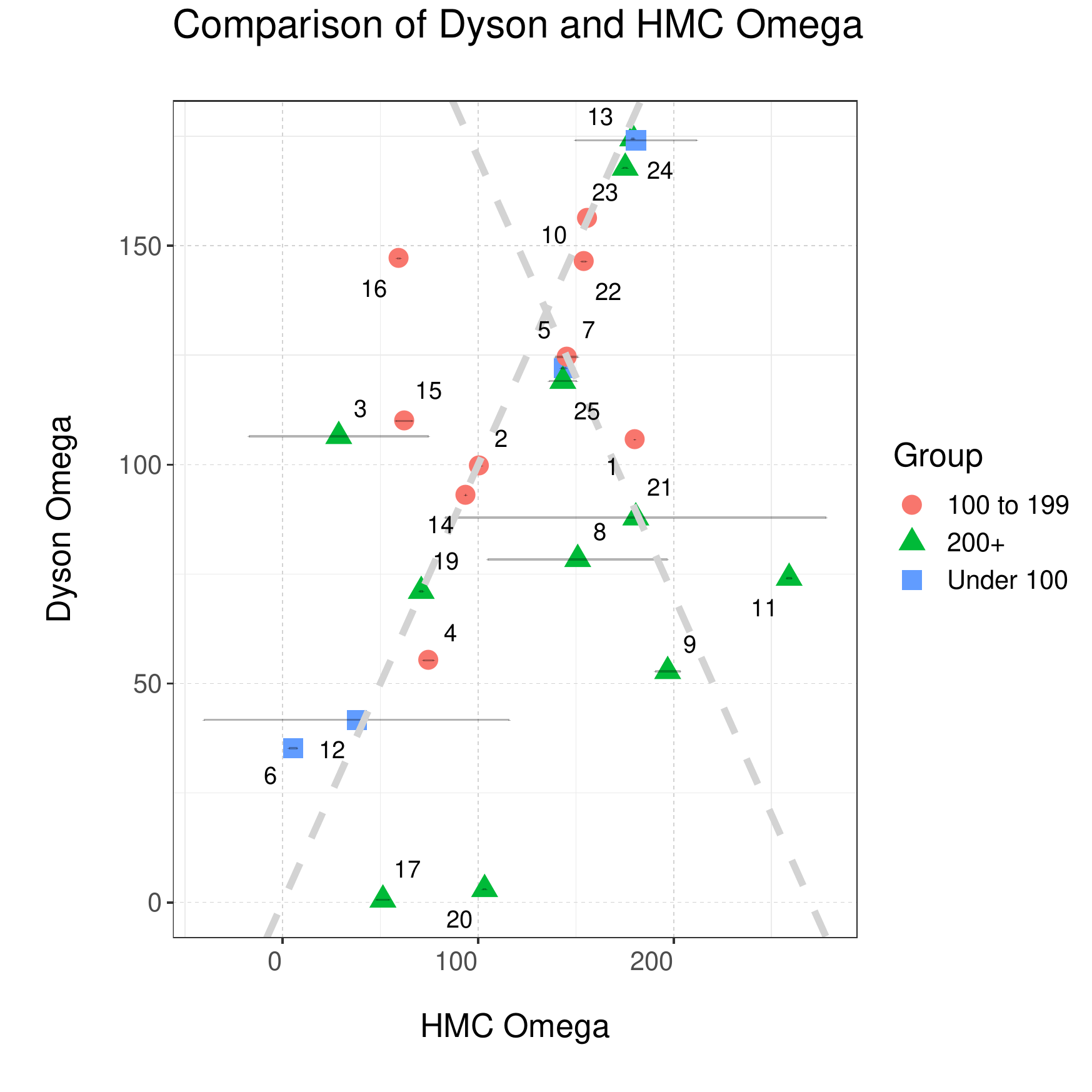} 
   \label{fig:Dyson_big_Omega}}
\end{subfloat}\hfil 

\caption{Comparison between HMC and Dyson optimal parameter estimates. Systems are denoted by their Dyson number. The dotted lines are those of perfect agreement between the two sets of parameter values. Observations are grouped by the HMC period estimates into less than 100 years, 100 to less than 200 years, and greater than or equal to 200 years. Dyson did not calculate uncertainties, so these are only plotted for the HMC estimates.
\label{fig:hmc_dyson_comp}}
\end{figure}

\subsection{Comparison with WinFitter Results}

Figure~\ref{fig:hmc_wf_comp} {\color{blue} (page~\pageref{fig:hmc_wf_comp})} plots the final parameter estimated from WinFitter \citep{Rhodes_2023} against this paper's MCMC results. Agreement is overall good, although we note that the longer period systems tend to have less well-constrained estimates for the parameters, as might be expected.  Coefficients of Determination ($R^2$) between the two sets of estimates, for each parameters, are shown in the sub-figures.  The regression slopes are not substantially different from unity, indicating good agreement between the two sets. Such agreement is comforting, and lends support to being able to later use the HMC-based program and technique on other systems which have not been modeled before --- indeed, a key driver for this project was the need to verify that a program written to model the orbit of V410 Puppis was correct \citep{Erdem_2022}, hence modeling the `known' systems of Dyson. The grey shaded regions give the formal two-sigma uncertainties in the regressions, showing that in general there is not a difference from a slope of perfect agreement for most of the parameters at the 95\% statistical confidence level. The exception is for period, where the WinFitter solutions for the longest period systems are in general smaller than the MCMC-based estimates. It is worth noting that the uncertainties given for these systems by the HMC method are large, indicating a lack of confidence in the point estimates. Overall the estimated uncertainties from the HMC method are larger than those from WinFitter, but it varies by parameter. Regression of the logarithm of the uncertainties gave the following relationships for the errors: $\log{P} = (0.41 \pm 0.10) + (0.94 \pm 0.12) \log{P_{W}}$ with $R^2 = 0.72$, $\log{a} =  (0.77 \pm 0.12)  \log{a_{W}}$ with $R^2 = 0.97$, $\log{e} =  (0.74 \pm 0.04) \log{a_{W}}$ with $R^2 = 0.95$, $\log{i} = (0.37 \pm 0.06) + (0.76 \pm 0.11) \log{i_{W}}$ with $R^2 = 0.69$, $\log{\omega} =  (1.09 \pm 0.0.33) \log{\omega_{W}}$ with $R^2 = 0.32$, and $\log{\Omega} = (0.24 \pm 0.09) + (0.66 \pm 0.14) \log{\Omega_{W}}$ with $R^2 = 0.52$ where the subscript $W$ refers to the results from the WinFitter fits by \cite{Rhodes_2023}.   In these equations, for simplicity we have used the parameter symbol as a placeholder for the error estimate of a parameter. We recommend the HMC approach as more rigorous, but given the substantially lower time required by WinFitter (seconds as opposed to MCMC runs which may take a day or more on a M1 Macbook Pro) these empirical scaling rules could be helpful for interpreting first looks using WinFitter. \cite{Rhodes_2023} had assumed a constant 5\% `mean observational error', which appears to be an underestimate of the actual scatter when compared with the $\sigma$ values given in Table~\ref{tab:dyson_parameters}.  It would be an interesting extension to this project to run the same fitting software (WinFitter) on these systems with the noise levels set to the values of $\sigma$ found by the MCMC fitting of the current paper, and see if these error estimates by WinFitter and our HMC method are in closer agreement.


\subsection{Comparison with Dyson Results}

Figure~\ref{fig:hmc_dyson_comp} (page~\pageref{fig:hmc_dyson_comp}) compares the optimal parameter estimates from \cite{Dyson_1921} and the HMC method of the current paper.  Naturally, the HMC method had access to an extra century for further observations, which helped constrain the orbital parameter estimates further.  The charts show overall good agreement between the Dyson estimates and those of this paper.  The longer period systems can show weaker agreement than for the shorter period ones, such as shown in Figure~\ref{fig:Dyson_P} where systems 8, 9, and 21 have clearly different estimated periods. Removing these three systems gives a regression slope of $0.99 \pm 0.08$, assuming a zero intercept, and the coefficient of determination $R^2 = 0.88$ which confirm good agreement.  Similarly, the agreement is good for $a$ (slope $0.96 \pm 0.04$, $R^2 = 0.95$, Figure~\ref{fig:Dyson_a}) and $e$ (slope $0.93 \pm 0.07$, $R^2 = 0.90$, Figure~\ref{fig:Dyson_e}). It is interesting to note that in general for systems with orbits of $ 100 <= P < 200 $ years, Dyson's estimated ellipticities appear systematically higher than those from the HMC approach. Matters become more complicated for the remaining parameters as there are ambiguities, e.g., inclination estimates can be symmetric around 90 degrees since from astrometric measurements alone we do not know the actual orbital direction of the star. Two dotted lines are therefore shown in each of Figures~\ref{fig:Dyson_omega}, \ref{fig:Dyson_i}, and \ref{fig:Dyson_big_Omega} to reflect this and show again good general agreement. While it is possible to `reflect' some of the results to allow calculation of linear correlations, we have chosen to leave the data unadjusted given some uncertainty which estimates should be reflected for some systems and we do not wish to present overly optimistic correlations through biased choices.

\begin{figure}
    \centering 
\begin{subfloat}[$P$]{
   \includegraphics[width=0.35\linewidth]{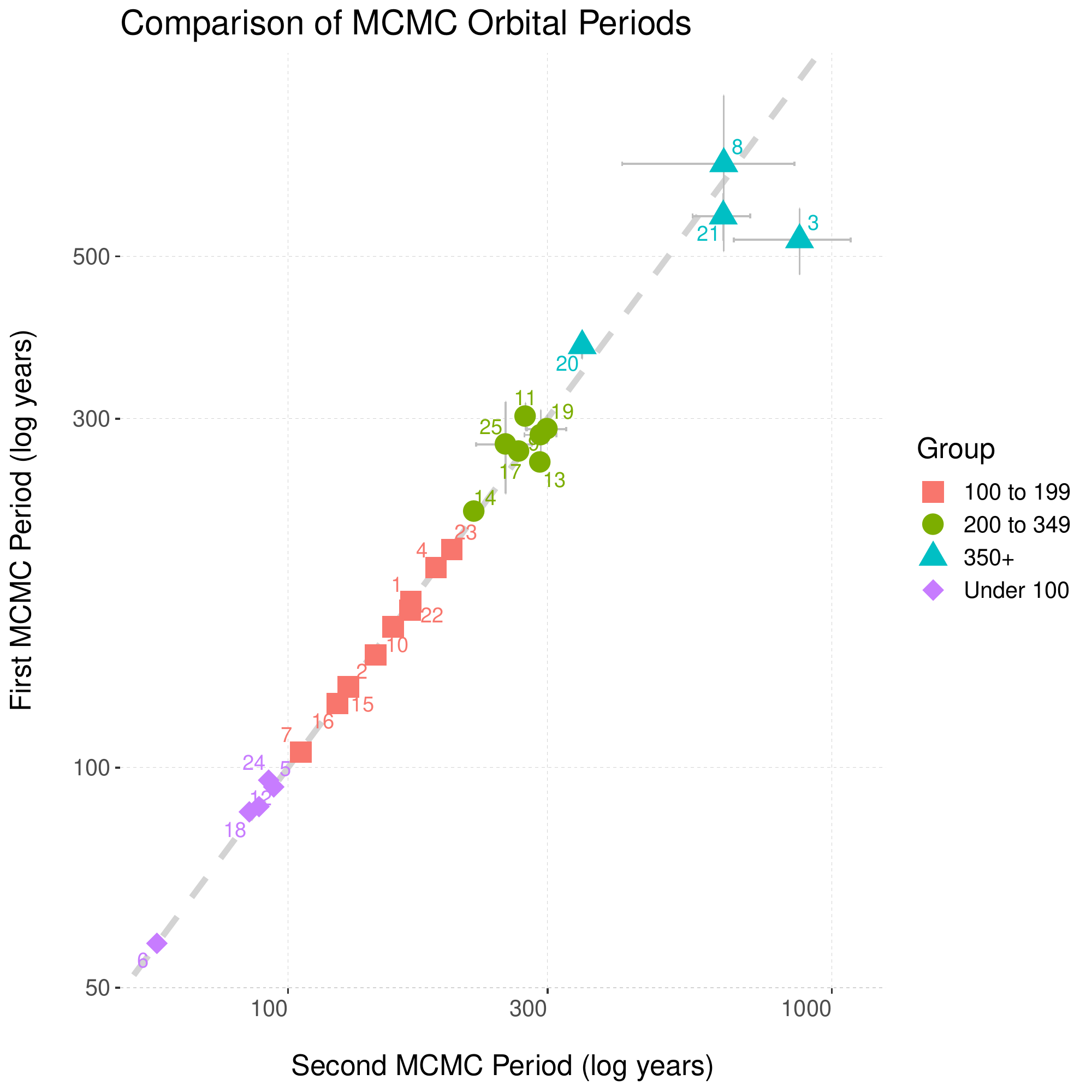} 
   \label{fig:2nd_Period_comp}}
\end{subfloat}\hfil 
\begin{subfloat}[$a$]{
   \includegraphics[width=0.35\linewidth]{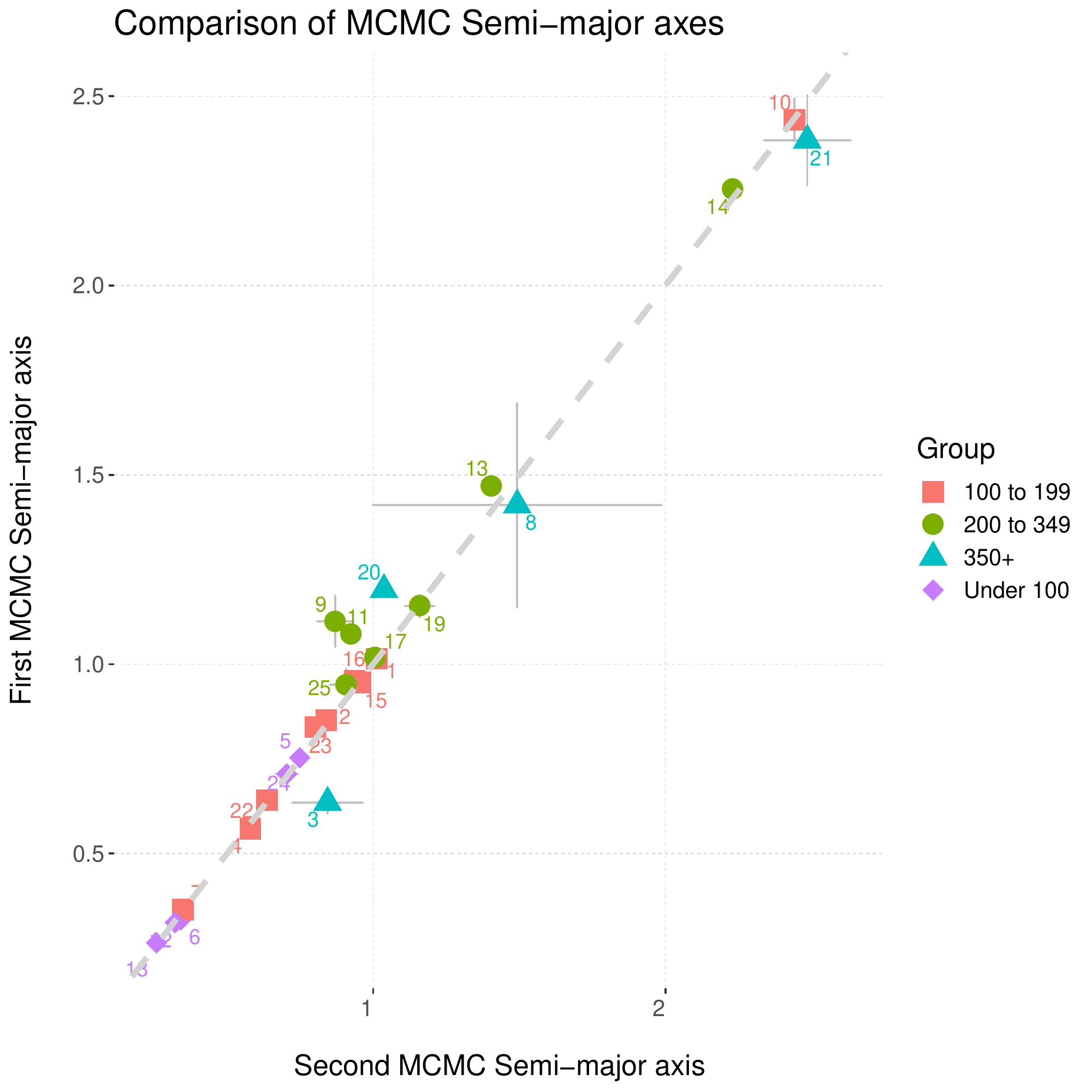} 
   \label{fig:2nd_axis_comp}}
\end{subfloat}\hfil 
\begin{subfloat}[$e$]{
   \includegraphics[width=0.35\linewidth]{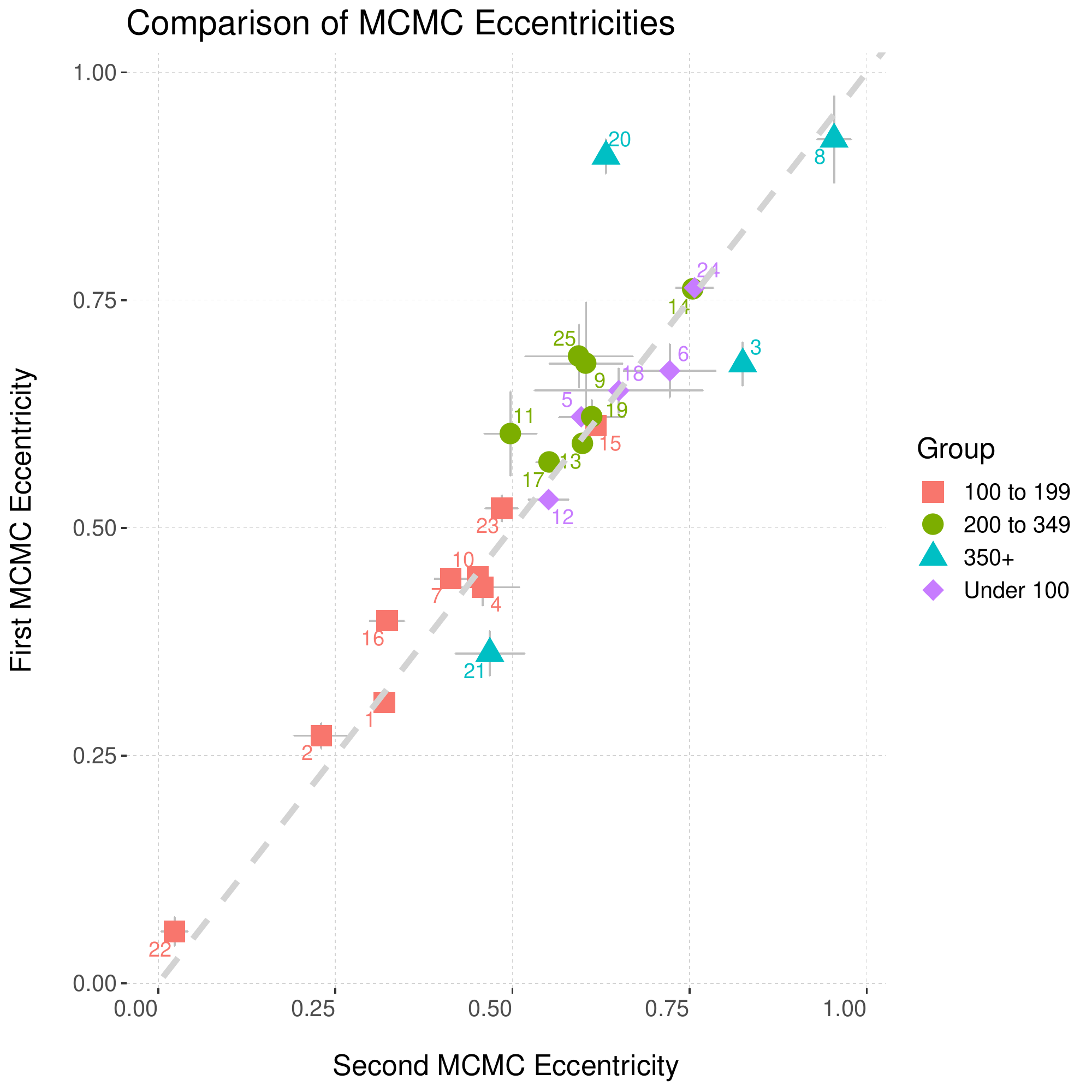} 
   \label{fig:2nd_ellip_comp}}
\end{subfloat}\hfil 
\begin{subfloat}[$\omega$]{
   \includegraphics[width=0.35\linewidth]{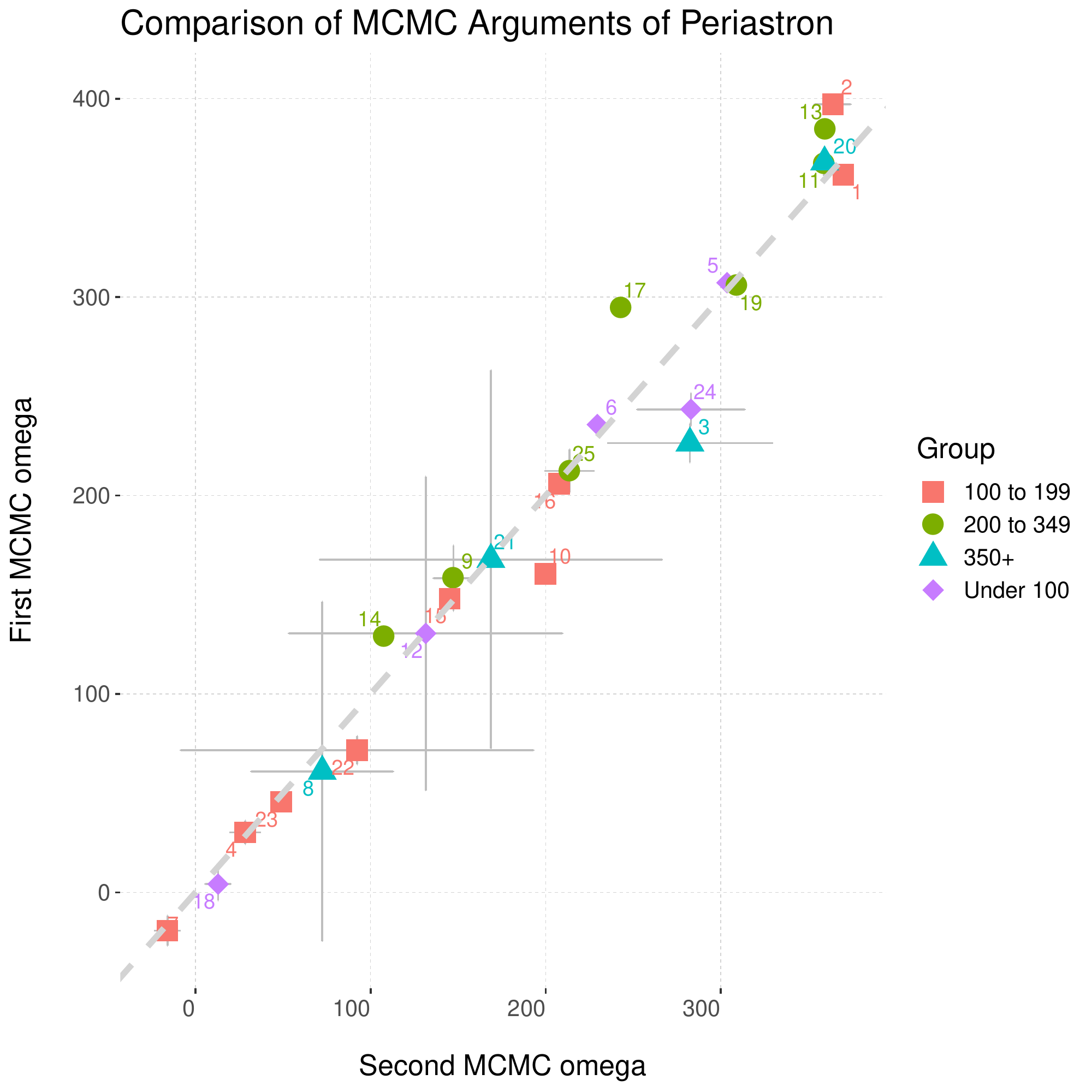} 
   \label{fig:2nd_little_omega_comp}}
\end{subfloat}\hfil 
\begin{subfloat}[$i$]{
   \includegraphics[width=0.35\linewidth]{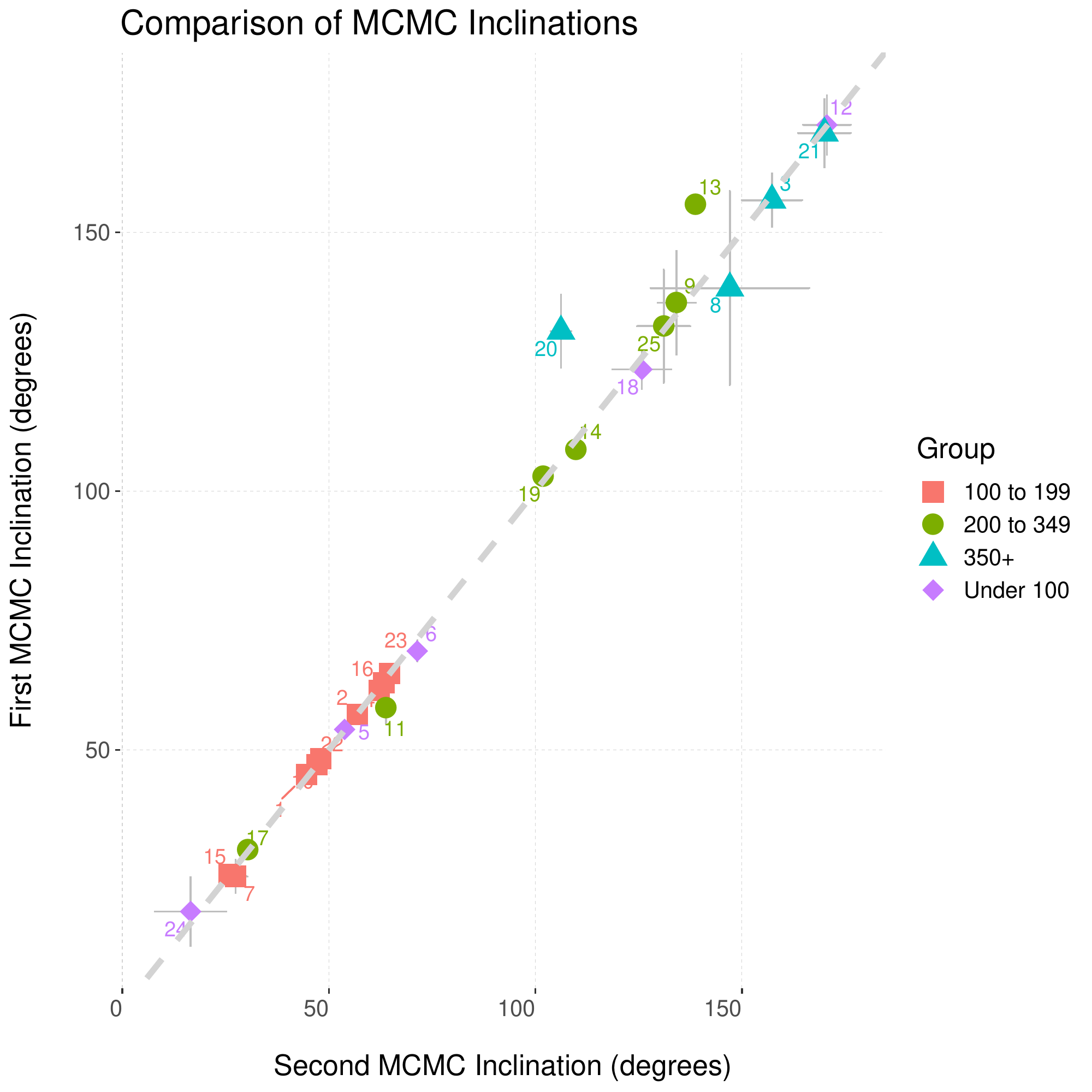} 
   \label{fig:2nd_inclination_comp}}
\end{subfloat}\hfil 
\begin{subfloat}[$\Omega$]{
   \includegraphics[width=0.35\linewidth]{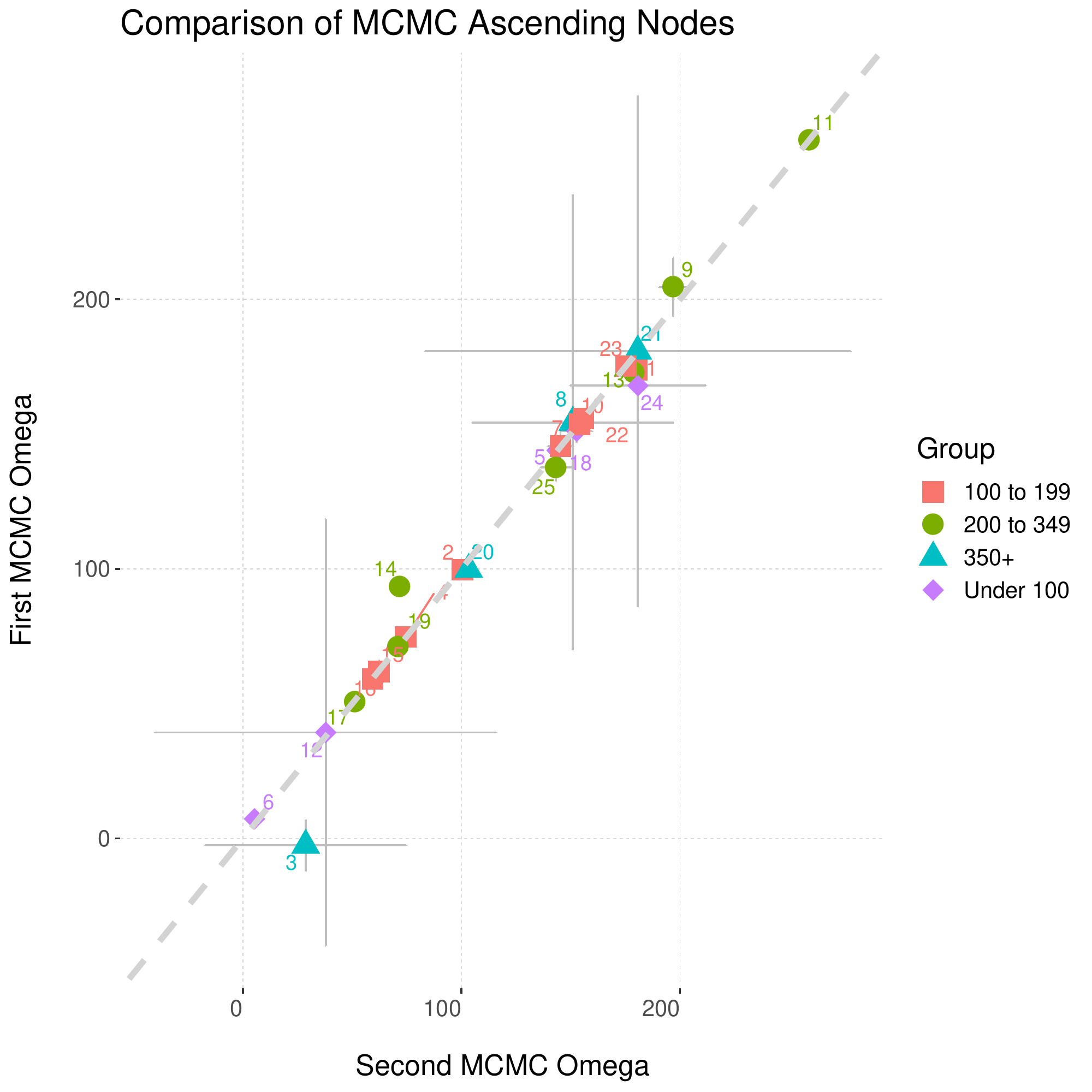} 
   \label{fig:2nd_big_omega_comp}}
\end{subfloat}\hfil 

\caption{Comparison between the MCMC fits with $\delta x$ and $\delta y$ included as free variables (``First MCMC", the $y$ axis) and those with the two parameters removed and $(x,y)$ set to $(0,0)$ (``Second MCMC'', $x$-axis). Systems are denoted by their Dyson number. The dotted lines are those of perfect agreement. Points are coded by color and shape by the estimated orbital periods (under 100 years, 100 to 199 years, 200 to 349 years, and greater than 350 years). Error bars correspond to the one standard deviation. Note that the orbital periods are given in log years.
\label{fig:mcmc_comp}}
\end{figure}

\begin{sidewaystable}
    \caption{ {\bf Parameter estimates from the second MCMC fitting} to the Dyson systems, numbered by appearance in Dyson (e.g., 1 refers to Dyson-1 or D1). Parameter columns are the same as in table~\ref{tab:dyson_parameters} (on page \pageref{tab:dyson_parameters}). Text colored blue indicates $3 \sigma$ differences with the results of the initial MCMC modeling given in Table~\ref{tab:dyson_parameters}. A high $\sigma$ was chosen to reduce the number of false positives that will occur in repeated statistical testing like this, to clearly show the systems with differences. System masses (in solar units) are calculated using the {\em Hipparcos} and {\em Gaia} parallaxes (given in the columns labelled `Hipparcos' and `Gaia'). }
    \centering
    \hspace{-2.5cm}
    {\footnotesize
    \begin{tabular}{||l|r|r|r|r|r|r|r|r|r|r||}
    \hline
System & $P$                & $a$                 & $e$               & $\omega$          & $i$             & $\Omega$            & Epoch              & $\sigma$               & Hipparcos   & Gaia \\
    \hline
1      & $168.56 \pm 0.57$  & $1.014 \pm 0.005$ & $0.308 \pm 0.003$   & {\color{blue} $1.6 \pm 1.2$}     & $45.2 \pm 0.5$  & {\color{blue} $173.93 \pm 0.72$}   & $1956.28 \pm 0.31$ & $0.0961 \pm 0.0025$ & $2.01 \pm 0.04$ & \\
2      & $142.51 \pm 1.63$  & $0.852 \pm 0.018$ & $0.272 \pm 0.014$   & $37.2 \pm 4.5$    & $61.5 \pm 1.5$  & $99.72 \pm 1.66$    & {\color{blue} $1900.64 \pm 1.42$} & $0.2244 \pm 0.0084$ & $1.89 \pm 0.06$ & \\
3      & {\color{blue} $526.67 \pm 54.22$} & $0.635 \pm 0.029$ & {\color{blue} $0.680 \pm 0.024$}   & $226.4 \pm 9.8$   & $156.2 \pm 5.1$ & $-2.56 \pm 9.67$    & $1911.40 \pm 0.71$ & $0.0867 \pm 0.0035$ & $2.10 \pm 0.35$ & \\
4      & $187.63 \pm 3.84$  & $0.565 \pm 0.011$ & $0.435 \pm 0.021$   & $30.2 \pm 6.0$    & $56.7 \pm 1.6$  & $74.52 \pm 2.50$    & $1887.34 \pm 1.87$ & $0.0790 \pm 0.0042$ & $5.37 \pm 0.30$ & \\
5      & $94.08 \pm 0.36$   & $0.753 \pm 0.012$ & $0.622 \pm 0.010$   & $307.2 \pm 1.6$   & $54.0 \pm 1.6$  & $143.94 \pm 1.47$   & $1887.88 \pm 0.39$ & $0.0830 \pm 0.0042$ & $3.25 \pm 0.12$ & $4.14 \pm 0.05$ \\
6      & $57.49 \pm 0.45$   & $0.324 \pm 0.014$ & $0.672 \pm 0.029$   & $235.7 \pm 3.7$   & $69.1 \pm 2.0$  & $7.28 \pm 2.17$     & $1943.09 \pm 0.42$ & $0.0710 \pm 0.0040$ & $2.46 \pm 0.11$ & \\
7      & $105.01 \pm 0.59$  & $0.352 \pm 0.005$ & $0.444 \pm 0.009$   & $-19.3 \pm 7.6$   & $25.6 \pm 3.2$  & $145.59 \pm 6.92$   & $1987.46 \pm 0.34$ & $0.0628 \pm 0.0026$ & $14.96 \pm 0.16$ & \\
8      & $669.28 \pm 160.97$& $ 1.420 \pm 0.269$& $0.927 \pm 0.048$   & $61.0 \pm 85.6$   & $139.2 \pm 18.7$& $154.31 \pm 84.59$  & $1893.64 \pm 3.12$ & $0.1064 \pm 0.0037$ & $4.70 \pm 0.48$ & $ 3.89 \pm 0.69$ \\
9      & $285.07 \pm 23.37$ & $1.113 \pm 0.068$ & $0.680 \pm 0.068$   & $158.4 \pm 16.5$  & $136.4 \pm 10.0$& $204.51 \pm 10.88$  & $1857.78 \pm 4.03$ & $0.1517 \pm 0.0050$ & $3.83 \pm 0.17$ & $ 2.92 \pm 0.15$ \\
10     & $155.79 \pm 0.35$  & $2.438 \pm 0.057$ & $0.446 \pm 0.002$   & {\color{blue} $160.6 \pm 0.6$}   & $47.2 \pm 0.3$  & $155.73 \pm 0.36$   & $1916.62 \pm 0.11$ & $0.1307 \pm 0.0031$ & $1.44 \pm 0.04$ & \\
11     & {\color{blue} $302.08 \pm 12.96$} & {\color{blue} $1.079 \pm 0.023$} & $0.603 \pm 0.046$   & $367.4 \pm 6.2$   & $58.1 \pm 2.9$  & $259.04 \pm 2.68$   & $1864.90 \pm 3.79$ & $0.0960 \pm 0.0040$ & $1.11 \pm 0.12$ & $1.16 \pm 0.07$ \\
12     & $88.43 \pm 0.45$   & $0.318 \pm 0.005$ & $0.531 \pm 0.010$   & $130.5 \pm 79.1$  & $170.7 \pm 5.7$ & $129.30 \pm 79.13$  & $1883.56 \pm 0.47$ & $0.0684 \pm 0.0029$ & $2.41 \pm 0.12$ & \\
13     & {\color{blue} $261.54 \pm 2.65$}  & {\color{blue} $1.470 \pm 0.008$} & $0.592 \pm 0.004$   & {\color{blue} $384.7 \pm 1.1$}  & {\color{blue} $155.3 \pm 1.1$} & {\color{blue} $172.70 \pm 0.96$}   & {\color{blue} $1864.31 \pm 0.32$} & $0.1150 \pm 0.0030$ & $1.94 \pm 0.05$ & $2.47 \pm 0.05$ \\
14     & $224.05 \pm 2.64$  & $2.254 \pm 0.019$ & $0.762 \pm 0.004$   & {\color{blue} $128.9 \pm 0.5$}   & $108.1 \pm 0.2$ & {\color{blue} $93.35 \pm 0.34$}   & {\color{blue} $1921.08 \pm 0.16$} & $0.1198 \pm 0.0037$ & $1.73 \pm 0.03$ & \\
15     & $128.91 \pm 0.53$  & $0.952 \pm 0.007$ & $0.612 \pm 0.005$   & $148.0 \pm 4.4$   & $25.9 \pm 1.8$  & $61.93 \pm 4.19$    & $1939.50 \pm 0.20$ & $0.1426 \pm 0.0036$ & $7.75 \pm 0.05$ & \\
16     & $122.36 \pm 0.83$  & $0.956 \pm 0.012$ & $0.398 \pm 0.010$   & $205.9 \pm 2.5$   & $63.0 \pm 1.0$  & $59.20 \pm 1.03$    & $1894.24 \pm 0.72$ & $0.1379 \pm 0.0057$ & $1.15 \pm 0.08$ & \\
17     & $270.66 \pm 5.12$  & $1.017 \pm 0.011$ & $0.572 \pm 0.008$   & {\color{blue} $294.6 \pm 3.4$}   & $30.7 \pm 1.7$  & $50.57 \pm 3.17$    & $1896.33 \pm 0.34$ & $0.1158 \pm 0.0039$ & $2.86 \pm 0.06$ & $3.69 \pm 0.04$ \\
18     & $86.97 \pm 1.19$   & $0.264 \pm 0.008$ & $0.651 \pm 0.025$   & $4.1 \pm 8.2$     & $123.5 \pm 3.7$ & $151.10 \pm 4.24$   & $2054.76 \pm 1.63$ & $0.1768 \pm 0.0146$ & $1.57 \pm 0.12$ & \\
19     & $290.23 \pm 13.31$ & $1.154 \pm 0.029$ & $0.622 \pm 0.018$   & $305.9 \pm 2.0$   & $102.9 \pm 0.6$ & $71.02 \pm 1.00$    & $1914.13 \pm 0.81$ & $0.1832 \pm 0.0065$ & $3.00 \pm 0.10$ & \\
20     & {\color{blue} $377.31 \pm 15.01$} & {\color{blue} $1.197 \pm 0.003$} & {\color{blue} $0.908 \pm 0.019$}   & $368.1 \pm 3.8$   & {\color{blue} $130.9 \pm 7.0$} & $99.75 \pm 2.70$    & $1886.48 \pm 1.66$ & $0.1101 \pm 0.0037$ & $3.23 \pm 0.15$ & $2.77 \pm 0.07$ \\
21     & $567.53 \pm 42.83$ & $2.384 \pm 0.119$ & $0.362 \pm 0.024$   & $167.8 \pm 95.4$  & $169.1 \pm 6.5$ & $180.74 \pm 94.94$  & $1866.16 \pm 3.82$ & $0.2103 \pm 0.0068$ & $5.44 \pm 0.14$ & $4.45 \pm 0.16$ \\
22     & $164.14 \pm 2.58$  & $0.641 \pm 0.06$  & $0.057 \pm 0.016$   & $71.7 \pm 7.1$    & $48.3 \pm 1.3$  & $153.62 \pm 1.19$   & $1873.57 \pm 3.17$ & $0.0725 \pm 0.0028$ & $2.52 \pm 0.08$ & \\
23     & $198.56 \pm 2.54$  & $0.824 \pm 0.011$ & $0.521 \pm 0.015$   & $45.7 \pm 2.5$    & $64.8 \pm 0.8$  & $175.19 \pm 1.15$   & $1896.37 \pm 0.82$ & $0.0830 \pm 0.0038$ & $3.29 \pm 0.07$ & $3.27 \pm 0.04$ \\
24     & $96.09 \pm 0.40$   & $0.710 \pm 0.018$ & $0.764 \pm 0.008$   & $243.4 \pm 8.3$   & $ 18.8 \pm 6.6$ & $168.00 \pm 8.29$   & $1905.42 \pm 0.24$ & $0.1063 \pm 0.0050$ & $1.28 \pm 0.10$ & \\
25     & $276.56 \pm 39.81$ & $0.946 \pm 0.041$ & $0.688 \pm 0.35$    & $212.4 \pm 10.9$  & $131.9 \pm 10.9$& $137.61 \pm 5.33$   & $1904.85 \pm 1.14$ & $0.0734 \pm 0.0045$ & $1.35 \pm 0.23$ & $1.38 \pm 0.23$ \\
       \hline 
    \end{tabular} 
    }
    \label{tab:2nd_dyson_parameters}
\end{sidewaystable}

\subsection{Parameter Reduction}
\label{sec:parameter_reduction}
As noted above, in results from the MCMC fits the parameters $\delta x$ and $\delta y$ were not statistically different from zero. This is as expected, from the fact that those variables are actually redundant in the model.  We therefore reran the fittings without these parameters to see the change in the estimates of the other parameter values. Overall, agreement is good as can be seen in Figure~\ref{fig:mcmc_comp} on page~\pageref{fig:mcmc_comp}, which plots parameters estimates from both groups of MCMC fittings by system and by parameter.  The dotted diagonal lines are those of perfect agreement between the two methods, with data points frequently falling with error of these lines.  Longer period systems were harder to `pin down', having the larger absolute change in estimates (which was signalled by the larger estimated standard deviations). The formal errors for some systems appear to be under-estimates, such as the ellipticities for the longer period systems Dyson-3 (D3) and -20 (D20). Even if the errors are tripled to be $3 \sigma$, they will not overlap the dashed line of agreement between the two fittings. This study did not make a detailed investigation of the likely observational `errors' (i.e., there was no weighting applied to the data to reflect different observational accuracies), which likely contributes towards these underestimates.  We plan to investigate improved estimation and handling of observational errors in further optimizations, extending this preliminary study. 

 
\begin{figure}
    \centering 
\begin{subfloat}[Dyson-17]{
    \includegraphics[width=0.45\textwidth]{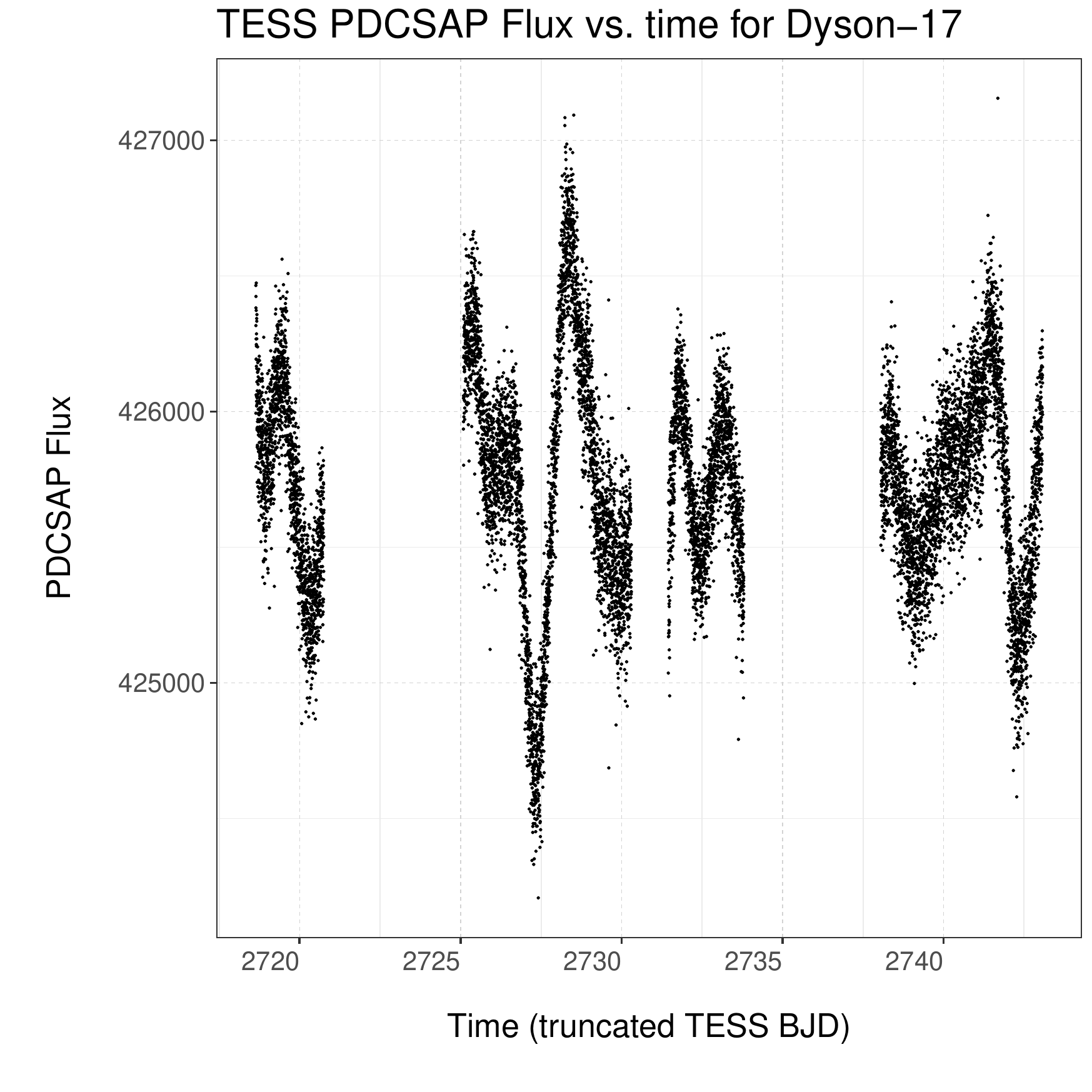}
    \label{fig:tess_d17}
} 
\end{subfloat}\hfil 
\begin{subfloat}[Dyson-20]{
    \includegraphics[width=0.45\textwidth]{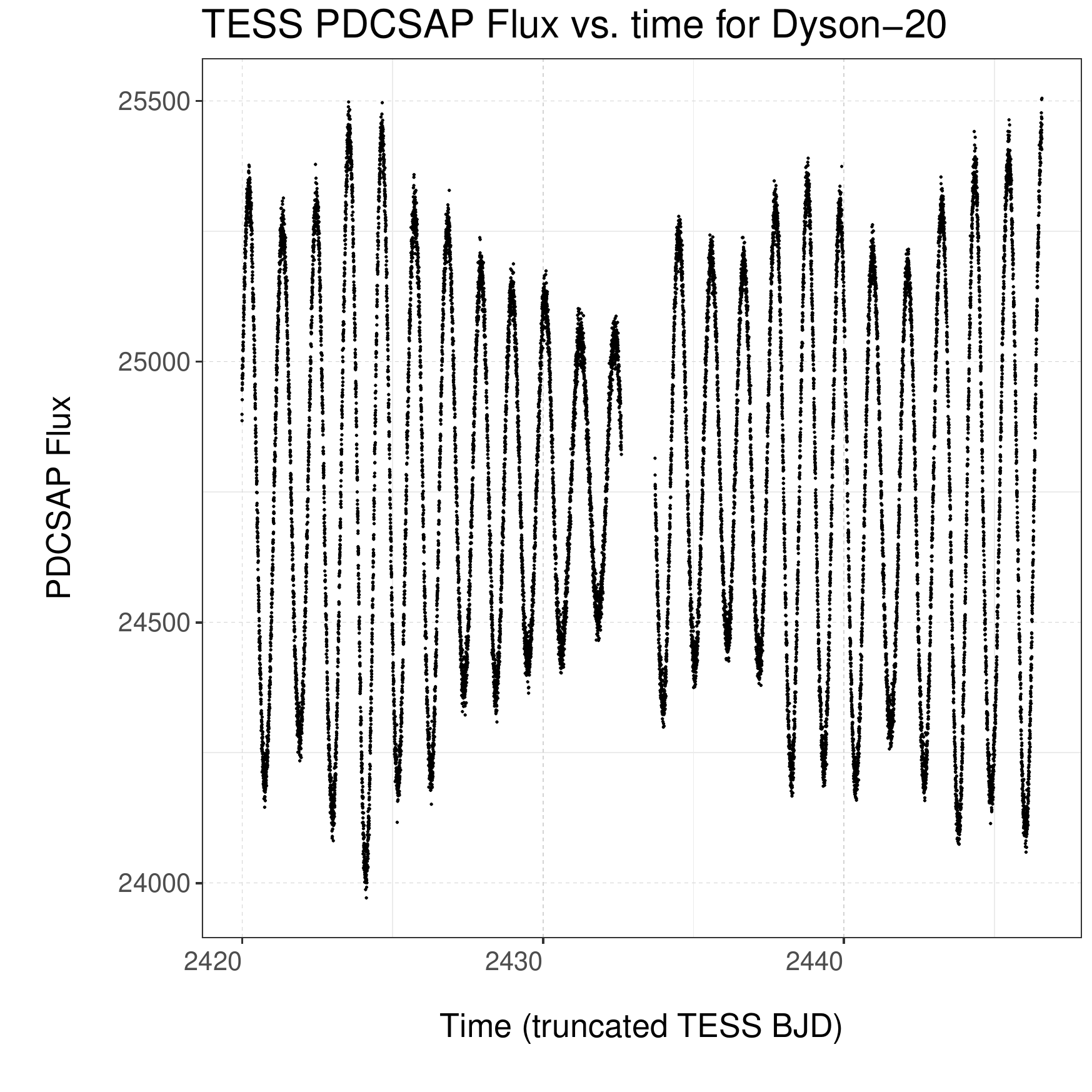}
    \label{fig:tess_d20}
}
\end{subfloat}\hfil 
 \caption{{\em TESS} photometry for two example systems (D17 and D20).  The fluxes are non-normalized Pre-search Data Conditioning Simple Aperture Photometry (PDC\_SAP) generated by the {\em TESS} team, which removed longstanding systematic trends.  Time is the Barycentric Julian Date (BJD) -- 2457000.
\label{fig:TESS_light_curves}}
\end{figure}

Table~\ref{tab:2nd_dyson_parameters} (page~\pageref{tab:2nd_dyson_parameters}) presents the results of these fits, where the blue-colored text indicates $3 \sigma$ differences between the results from the original MCMC fits (Table~\ref{tab:dyson_parameters}, page \pageref{tab:dyson_parameters}) and these with a fixed origin. The significant differences are mainly in the longer period systems D3, D11, D13, and D20 with agreement being good for the other systems. The data for D3 and D20 are sparse for roughly the first half of the observation periods, while those for D11 and D13 correspond to arcs without much curvature.  The parameter estimates for the other systems are overall within statistical error of those from the previous fits. Removing the four mentioned systems from regression analyses comparing the parameter estimates, we first found that intercept terms were not statistically significant. Regressions through the origin for ($\Omega, \omega, e, i, a, P$) had coefficients of determination ($R^2$) of $(0.9979, 0.9805, 0.9940, 0.996, 0.9973, 0.9956$) respectively, with slopes of (1.0004, 0.9859, 0.9844, 1.008, 0.998, 0.996) and corresponding standard errors of (0.010, 0.021, 0.017, 0.005, 0.012, 0.015).  Estimates for the dynamical masses are also given in Table~\ref{tab:2nd_dyson_parameters}. Taking log masses (from the two sets of MCMC fits) gave regression slopes of $1.0 \pm 0.1$ for the systems with {\em Gaia} distances and $0.99 \pm 0.05$ for {\em Hipparcos} distances.

Inclusion of the parameters $\delta x$ and $\delta y$ is not needed for this kind of study (they act as `nuisance' parameters), and we will not include them in subsequent similar studies given this comparison. Additionally, removal of the parameters reduces by 2 the number of dimensions being searched in the optimization, lowering the computational load. The results given in the current section and Table~\ref{tab:2nd_dyson_parameters} are this paper's final results for the analyzed systems.

\newpage
\section{Discussion}

This paper has presented MCMC analyses of astrometric data for 25 systems, updating orbital parameters estimates given by \cite{Dyson_1921} a century ago.  It has shown reasonable agreement between the two studies, and an even stronger agreement with an earlier investigation by the current team \citep{Rhodes_2023} using a tool (in WinFitter) they developed employing point estimation supplemented with examination of the Hessian matrix to estimate uncertainties in the parameters.  The HMC method led to larger estimates than those calculated by WinFitter, which we believe are more robust being based on exploration of the Bayesian posterior distribution. Finally, we include parallax estimates from the {\em Hipparcos} and {\em Gaia} missions to present estimates and associated uncertainties of the dynamical masses of the systems.

Analysis of double star orbits is a useful tool in the variable star analyst's toolkit, and can be supplemented with other techniques to better improve understanding of the system. For instance, \cite{Mendez_2017} extend their MCMC astrometric analysis with the inclusion of radial velocities, which help to resolve issues with the direction of the orbital movement and the ambiguities noted above for some of the optimized parameters. Member stars of a double system can be variable, indeed cursory examination of the {\em TESS} space telescope \citep{Ricker_2014} photometry for the Dyson systems (see Figure~\ref{fig:TESS_light_curves}) indicates possible variable for a number of systems (D3, D5, D9, D10, D13, D17, D18, D20, D22, \& D23).  Analysis of such variability, such as through asteroseismology studies (see, e.g., \citealt{Aerts_2010}), could provide additional insights such as mass estimates. It is also increasingly common for systems to be recognised as multiple systems, such as with the case of V410 Puppis \citep{Erdem_2022} where astrometric analysis of the orbit of third system member about an inner eclipsing binary pair helped provide insights into the overall system. 

Observations and analysis of visual binaries is not time consuming (see, e.g., \citealt{Cleveland_2022}) in the actual data collection, nor is the measurement of positions and angles (say from CCD images).  While the `payback' of such observations may not be immediate, we end with the thoughts of Hertzsprung (as given in \citealt{Mason_2006}):
\begin{quote}
{\em ``The debt to our ancestors for the observations they made to our benefit, we can pay only by doing the same for our descendants."}
\end{quote}
Using such astrometric data collected by previous generations of astronomers led to feelings of connection with both them and the development of astronomy with time, as well appreciation for the work by our predecessors.  While missions such as {\em Gaia} will add data for astrometric binaries, on-going measurements (for instance in periods outside such surveys) will no doubt be appreciated by astronomers in the future. 

Further details and background on this project may be found in \cite{Soh_2023}.


\begin{acknowledgments}

This work has made use of data from the European Space Agency (ESA) mission {\it Gaia} (\url{https://www.cosmos.esa.int/gaia}), processed by the {\it Gaia} Data Processing and Analysis Consortium (DPAC, \url{https://www.cosmos.esa.int/web/gaia/dpac/consortium}). Funding for the DPAC has been provided by national institutions, in particular the institutions participating in the {\it Gaia} Multilateral Agreement. We thank the University of Queensland for collaboration software.  This paper includes data collected with the TESS mission, obtained from the MAST data archive at the Space Telescope Science Institute (STScI). Funding for the TESS mission is provided by the NASA Explorer Program. STScI is operated by the Association of Universities for Research in Astronomy, Inc., under NASA contract NAS 5–26555. This research has made use of the Washington Double Star Catalog maintained at the U.S. Naval Observatory (USNO). We thank the USNO and Dr.\ Rachel Matson for access to the WDS data. We thank the anonymous referee for their comments and guidance which improved the paper.

\end{acknowledgments}


\allauthors
\end{document}